\documentclass[dvips]{class}
\usepackage{supertabular,lscape,epsfig}
\usepackage{amssymb}
\usepackage{amsmath}
\usepackage{mathptmx}
\usepackage{booktabs}
\usepackage{rotating}
\usepackage{multirow}
\usepackage[a4paper, textwidth=13.7cm, textheight=21cm,
            hcentering=true, vcentering=true]{geometry}

\begin{document}

\begin{Titlepage}

\Title{Doubling the Number of Blue Large-Amplitude Pulsators: Final Results of Searches for BLAPs in the OGLE Inner Galactic Bulge Fields}

\Author{J.~~B~o~r~o~w~i~c~z$^1$,~~P.~~P~i~e~t~r~u~k~o~w~i~c~z$^1$,~~J.~~S~k~o~w~r~o~n$^1$,~~M.~J.~~M~r~\'o~z$^1$,\\ A.~~U~d~a~l~s~k~i$^1$,~~M.~K.~~S~z~y~m~a~\'n~s~k~i$^1$,~I.~~S~o~s~z~y~\'n~s~k~i$^1$,~~K.~~U~l~a~c~z~y~k$^2$,\\ R.~~P~o~l~e~s~k~i$^1$,~~S.~~K~o~z~{\l}~o~w~s~k~i$^1$,~~P.~~M~r~\'o~z$^1$,\\ D.~M.~~S~k~o~w~r~o~n$^1$,~~K.~~R~y~b~i~c~k~i$^{1}$,~~P.~~I~w~a~n~e~k$^1$,\\ M.~~W~r~o~n~a$^{1,3}$~~and~~~M.~~G~r~o~m~a~d~z~k~i${^1}$\\}{${}^1$ Astronomical Observatory, University of Warsaw, Al. Ujazdowskie 4, 00-478 Warszawa, Poland \\ ${}^2$ Department of Physics, University of Warwick, Coventry CV4 7AL, UK \\ ${}^3$ Department of Astrophysics and Planetary Science, Villanova University, 800 East Lancaster Avenue, Villanova, PA 19085, USA \\
}


\footnotesize{October, 2025}
\end{Titlepage}

\Abstract{Blue Large-Amplitude Pulsators (BLAPs) are rare short-period ($\lesssim$80~min) pulsating variable stars exhibiting large-amplitude brightness variations (typically between 0.1 and 0.4 mag). As a recently discovered class of radial-mode pulsators, the origin and nature of these variables remain the subject of ongoing investigations. Here, we present a comprehensive summary of all BLAPs identified in the data of the Optical Gravitational Lensing Experiment (OGLE), including the discovery of 87 new BLAPs in the inner Galactic bulge fields. We performed a systematic search for periodic signals in the \textit{I}-band light curves of more than 400 million stars with magnitudes down to $I = 21$. Our search effectively doubles the number of these variables to almost 200. The detected BLAPs exhibit pulsation periods between roughly 5 and 76 minutes. The analyzed dataset covers a timespan from 2001 to 2024, with some stars observed up to 20,000 times, providing the temporal coverage needed to study period and amplitude variations. We report on three objects that show enormous period changes, at a rate of $10^{-5}$~yr$^{-1}$, which could provide important clues to the evolutionary status of BLAPs. Full dataset is incorporated into the publicly available OGLE Collection of Variable Stars (OCVS), enabling future studies of these enigmatic objects.}

{Stars: oscillations (including pulsations) --- Stars: variables: Blue Large-Amplitude Pulsators}

\section{Introduction}

Pulsating variable stars have been recognized and studied for more than a century, serving as fundamental astrophysical laboratories that provide direct constraints on stellar structure and evolution. They populate nearly every part of the Hertzsprung-Russell (HR) diagram: from long-period variables and classical Cepheids varying with periods from days to months, through RR~Lyrae, $\delta$~Scuti, $\gamma$~Doradus, and $\beta$~Cephei-type stars pulsating with periods of hours to more compact objects like white dwarfs (WDs) and hot subdwarfs (sdBs) oscillating with periods of minutes. Many of these classes, such as Cepheids and RR~Lyrae stars, obey tight period-luminosity relations and serve as standard candles for distance measurements. As both precise distance benchmarks and probes of stellar evolution, these variables also provide vital information on the overall structure and history of the Galaxy. In recent decades, optical wide-field time-domain photometric ground-based surveys such as ASAS, ASAS-SN, OGLE, and ZTF as well as space missions like Kepler, Gaia, and TESS have dramatically increased the number of variable stars, uncovering thousands of new pulsating stars of all types and enabling precise tests of stellar evolution and interior physics through asteroseismology. This observational expansion has facilitated the identification of novel classes of rapid pulsators, among which the \textit{Blue Large-Amplitude Pulsators} (BLAPs) have emerged as a distinct and intriguing group.

Initially, BLAPs were identified as a by-product of searches for other types of variable stars. Their low apparent brightness (typically $I \gtrsim 15$~mag) and the fact that radial-mode pulsators were rather not expected blueward of the main sequence in the HR diagram meant that their discovery only became feasible with the advent of wide-field, time-domain photometric surveys. The prototype, OGLE-BLAP-001 ($P = 28$~min), was first found in the OGLE-III data and initially classified as a $\delta$~Scuti variable (Pietrukowicz \etal 2013). Subsequent spectroscopic studies, however, placed its effective temperature at around 30,000~K. Together with thirteen similar objects in the area of the Galactic bulge, these stars were identified as a new class of variables, named "Blue Large-Amplitude Pulsators" in virtue of their observed properties (Pietrukowicz \etal 2017).

The original sample of BLAPs suggested pulsation periods in the range of 20-40~min, with brightness variations of approximately 20--45\%. Derived atmospheric parameters like surface gravity $\log g \approx 4.2$--4.6 dex and helium abundance $\log ( N_{\mathrm{He}} / N_{\mathrm{H}} )$ $\sim -0.6$ dex, placed these objects close to the hot subdwarfs domain. The first extension of the class came in 2019, when four objects with much shorter periods, between 3 and 8 minutes, were discovered in the ZTF data (Kupfer \etal 2019). In the case of these stars, surface gravity values were found to be about an order of magnitude higher, and they were therefore dubbed "high-gravity BLAPs" (HG-BLAPs).

Further expansion of the known BLAP sample occurred soon thereafter. Based on combined data from the ZTF, Gaia, and Pan-STARRS surveys, McWhirter \& Lam (2022) identified over twenty BLAP candidates (named ZGP-BLAP-NN), including several objects located at high Galactic latitudes ($|b|>10$ deg). Around the same time, four additional BLAPs were confirmed in the OmegaWhite data (Ramsay \etal 2022). Another BLAP was independently identified in the SkyMapper survey (Cheng \etal 2024). Even with these additions, the total known sample remained modest, only a few dozen objects.

Due to the unusual properties of BLAPs: relatively high helium abundance, location in a sparsely populated region of the HR diagram above hot subdwarfs, and a characteristic pulsation behavior, their evolutionary status and formation mechanism have remained uncertain since their discovery. It was relatively quickly established that the classical $\kappa$-mechanism, driven by the "Z-bump" in the opacity curve, generates pulsations in BLAPs (Byrne \& Jeffery 2018, 2020). Radiative levitation and atomic diffusion are believed to be responsible for accumulating a sufficient amount of metals in the pulsation-driving layer. Over the years, three competing structural models of BLAPs have been proposed:
\begin{enumerate}
    \item Low-mass ($\sim$0.3\,$M_\odot$) pre-WDs with helium cores and residual hydrogen-burning shell (Romero \etal 2018),
    \item Helium-core burning post-giants of intermediate mass ($\sim$0.45\,$M_\odot$) evolving toward the extended horizontal branch (EHB, Wu \& Li 2018, Byrne \& Jeffery 2018),
    \item Post-EHB stars undergoing stable helium-shell burning above the CO core, typically with $M \gtrsim 0.45\,M_\odot$ (Xiong \etal 2022, Lin \etal 2023).
\end{enumerate}

Since BLAPs cannot be explained as products of single-star evolution, binary interactions are expected to play a central role in their formation. BLAPs have been proposed to be surviving companions of Type Ia supernovae (Meng \etal 2020), remnants of Roche-lobe overflow or common-envelope evolution (Byrne \etal 2021), or stellar merger products. In the latter case, possible progenitor systems include a main-sequence star with a white dwarf companion (Zhang \etal 2023). This highlights the importance of investigating binarity among observed BLAPs. Recently, it has been demonstrated that BLAPs may originate as merger products of two extremely low-mass white dwarfs (Kolaczek-Szymanski \etal 2024). This scenario gains further support from the discovery of two magnetic BLAPs (Pigulski \etal 2024), whose rotationally split, equidistant frequencies can be explained within the framework of the oblique pulsator model.

Binary BLAPs have also been directly identified. Using TESS data, the first such case was revealed in HD 133729, a pulsator with $P = 32.37$ min orbiting a B-type main-sequence star with a 23.08-day period (Pigulski \etal 2022). Detailed binary evolution modeling suggests that BLAPs such as HD 133729 can most plausibly form through the pre-WD Roche-lobe overflow channel, with non-conservative mass transfer playing a key role in reproducing the observed properties (Zhang \etal 2025a). Observations from Tsinghua University-Ma Huateng Telescopes for Survey (TMTS) suggest that star TMTS-BLAP-1 (=ZGP-BLAP-01) may reside in a wide binary with an orbital period of about 1576 days (Lin \etal 2023). More recently, Korea Microlensing Telescope Network (KMTNet) survey photometry indicated that OGLE-BLAP-006 may have two wide-orbit companions with periods of about 4700 and 6300 days, suggesting membership in a hierarchical triple system (Kim \etal 2025).

A detailed observational study of OGLE-BLAP-009 (Bradshaw \etal 2024) provided one of the first comprehensive constraints on the physical parameters of a BLAP, showing consistency with the low-mass ($\approx 0.3$M$_\odot$) helium-core pre-WD scenario. Linear pulsation grids for BLAPs, exploring a range of convection treatments and metallicities, have mapped their instability regions and established new theoretical period relations, with results most consistent with a low-mass origin (Das \etal 2024). Recent pulsation models for BLAPs (Jeffery 2025) have demonstrated that their instability strips, amplitudes, and mode behavior depend strongly on stellar mass, effective temperature, and envelope composition, offering a framework to interpret the observed diversity of light curve shapes. In a recent study, Zhang \etal (2025b) showed that helium-burning stars can contribute significantly to the Galactic BLAP population, predicting their numbers, binary properties, and detectability in future surveys. Non-adiabatic pulsation analyses have confirmed that the observed properties of BLAPs are consistent with radial fundamental modes, while their pulsation excitation is mainly driven by metal abundance rather than surface composition or envelope thickness (Wu \& Li 2025).

The discovery of new BLAPs in recent years has substantially expanded the known parameter space of the class. Systematic searches of OGLE-IV fields in the Galactic disk (Borowicz \etal 2023a) and the outer Galactic bulge (Borowicz \etal 2023b) uncovered 20 and 31 additional BLAPs, respectively, more than doubling the previously known sample. These new discoveries bridge the apparent gap between two subgroups of BLAPs (normal and high-gravity stars) that initially appeared distinct. In parallel, 23 additional BLAPs have been reported in the OGLE-IV inner Galactic bulge fields (Pietrukowicz \etal 2025a), which also demonstrated a dichotomy between helium-rich and helium-poor subgroups. That work revealed the presence of multi-mode pulsations, exemplified by OGLE-BLAP-030. Very recently, OGLE-BLAP-006 was also shown to exhibit multi-modality with two additional independent frequencies detected (Kim \etal 2025). 

In this study, we present results of extensive searches for new BLAPs in the inner Galactic bulge based on OGLE-IV data supplemented with archival OGLE-III observations. Section~2 provides a brief overview of the OGLE survey and the spectroscopic material used, while Section~3 outlines the search methodology. The new entries to the OGLE catalog of BLAPs with their basic properties are introduced in Section~4, followed in Section~5 by a description of several additional variables that may be of interest. Section~6 focuses on the analysis of period changes and the search for secondary frequencies among the reported objects. Considerations regarding the completeness and purity of the resulting catalog are discussed in Section~7, and the main findings of this work are summarized in Section~8.

\section{Observations}

Most of the photometric data analyzed in this study were obtained during the fourth phase of the Optical Gravitational Lensing Experiment (OGLE-IV), a long-term observational program that began in March 2010. OGLE has been a pioneering large-scale photometric survey for over three decades. The survey uses the 1.3-meter Warsaw telescope located at the Las Campanas Observatory, Chile, operated by the Carnegie Institution for Science. Initially focused on detecting microlensing events toward the Galactic bulge, OGLE has significantly expanded its scope over the years. During OGLE-IV, the project has monitored up to 3,600~deg$^2$ of the southern sky, including the Galactic bulge, the Galactic disk, and the Magellanic System. Observations are conducted using a 269-megapixel mosaic camera consisting of 32 CCD detectors, offering a 1.4~deg$^2$ field of view and a pixel size of 0.26 arcsec. Photometry is primarily carried out in the Cousins \textit{I}-band, with additional Johnson \textit{V}-band images to provide color information. Exposure times vary by region: typically 25~s for Galaxy Variability Survey (GVS) fields (disk and outer bulge), 100~s for inner bulge fields, 150~s for selected Sagittarius dwarf spheroidal galaxy and Magellanic System fields. The survey employs difference image analysis for the data reduction, optimized for crowded stellar fields. The OGLE-IV long-term, high-precision photometry has led to the discovery and classification of over one million variable objects, including transient, irregular, and periodic sources of various types. Further technical details on the OGLE-IV project can be found in Udalski et al. (2015).

In addition, a subset of the data used in this work originates from the third phase of the survey (OGLE-III), conducted between 2001 and 2009. The survey provided extensive coverage of the inner regions of the Galactic bulge, spanning nearly $69~\mathrm{deg}^2$. Observations were performed with an eight-detector mosaic camera offering a field of view of $0.35^\circ \times 0.35^\circ$ and a scale of 0.26 arcsec/pixel. Photometric monitoring was carried out primarily in the \textit{I}-band, with the number of measurements per star ranging from about 30 to more than 5000, depending on the field. Additional data were obtained in the \textit{V}-band, with both filters closely resembling the standard Johnson-Cousins photometric system, as in the case of OGLE-IV. More information on the OGLE-III survey, including its setup and technical details, is provided by Udalski \etal (2008).

Spectroscopic observations of OGLE-BLAP-098, OGLE-BLAP-161, and OGLE-BLAP-181 were carried out over two nights (26/27 and 27/28 July 2024) using the EFOSC2 spectrograph mounted on the 3.58-m New Technology Telescope (NTT) at the European Southern Observatory La Silla Observatory, Chile, under ESO program 113.26U3. The observations were performed with grism \#7, covering a wavelength range of 3270-5240~\AA, and a slit width of 1.0 arcsec. Atmospheric conditions were good, with moderate seeing of approximately 1.35 arcsec. For each target, two individual exposures were taken to enhance the signal-to-noise ratio and facilitate the removal of cosmic-ray events. The resulting spectra achieve a measured resolution of about 5~\AA~at 4000~\AA. Data reduction was carried out with the EFOSC2 pipeline (version 2.3.9), incorporating bias subtraction, flat-field correction, sky subtraction, wavelength calibration, and spectral extraction.

\section{Identification of BLAPs}

The search for variable objects in the inner Galactic bulge (BLG) fields was conducted in a manner similar to that described in Borowicz \etal (2023a, 2023b). This time, 114 BLG fields were analyzed, covering the most dense area of the bulge in the optical range. Previously, 13 objects were identified in these fields by Pietrukowicz \etal (2017), with 23 more reported in a subsequent study (Pietrukowicz \etal 2025a). In the present study, we performed a systematic and profound search for any short-period variable objects toward the inner bulge. Detailed information on the locations of the OGLE-IV fields is available at:  
\begin{center}
\textit{https://ogle.astrouw.edu.pl/sky/ogle4-fields.html}
\end{center}

Our search was carried out using the \texttt{FNPEAKS}\footnote[1]{http://helas.astro.uni.wroc.pl/deliverables.php?lang=en\&active=fnpeaks} code, which fits a sinusoidal signal to unevenly sampled data. Full light curves were examined over the frequency range 0--500~cycles per day (c/d). For each light curve, the dominant peak in the periodogram was recorded. In this way, more than 400 million OGLE-IV \textit{I}-band light curves were examined, spanning the magnitude range $12 < I < 21$, consistent with the typical photometric depth of the survey. Only objects with a significant signal-to-noise ratio were retained, where the threshold varied from field to field, since the number of observations in the inner BLG fields ranged from fewer than 100 (for the heavily obscured fields in the direction of the Galactic center) to more than 15,000 in the case of the OGLE-IV data (years 2010--2024). To analyze the short-period variability, the brightness measurement timestamps were converted from Heliocentric Julian Date (HJD) to the more accurate Barycentric Julian Date (BJD\textsubscript{TDB}). After removing evident outliers from the light curves, the previously determined periods were refined using the \texttt{TATRY} code (Schwarzenberg-Czerny 1996), which employs multi-harmonic fitting. Instrumental magnitudes were subsequently transformed to the standard photometric system. The calibration procedure, following the method described in Udalski \etal (2015), provides an accuracy of approximately 0.02~mag in the Johnson-Cousins system.

The next step was a visual inspection of the remaining light curves (after the signal-to-noise ratio cut) to remove spurious detections and artifacts mimicking genuine variables. In addition, some light curves required their detected periods to be adjusted (typically multiplied by two or three) to achieve proper phasing. For some of the fields with sufficiently dense time coverage, individual observing seasons (2011, 2013, and 2019) were also analyzed independently to search for potential objects exhibiting rapid period changes. After combining and cleaning all lists, more than 4000 genuine variable stars with periods shorter than 0.06~d were identified, including over 1000 objects with periods below one hour.   

Finally, in order to assess completeness and to compile a catalog covering the entire OGLE collection, 267 OGLE-III fields located in the Galactic bulge were also investigated, encompassing a total of 338 million stars. The same procedures as for the OGLE-IV dataset were applied. This led to the discovery of several additional BLAPs, mostly objects which in OGLE-IV fields are located in the gaps between individual CCD detectors. 

Subsequently, candidates for BLAPs were selected from the obtained variable star sample. The selection criteria were based on the light-curve morphology, period, amplitude, and color, following the photometric properties characteristic of spectroscopically confirmed BLAPs reported in the literature. We independently recovered all 36 BLAPs previously identified in the inner BLG fields (Pietrukowicz \etal 2017, 2025a). These stars are not reclassified here, since they are already part of the OGLE catalog.

\section{The Collection of BLAPs}

In this section, we present the collection of 90 previously unclassified BLAPs identified in the OGLE data. Eighty-seven of these objects are present in the literature as BLAPs for the first time, which increases the number of all known stars of this class to around 200. We note that since 2017, the number of known objects of this rare class has increased by more than an order of magnitude. The sample was constructed based on candidates selected through the procedures described above and refined by applying photometric criteria characteristic for spectroscopically confirmed BLAPs. For each object, we provide its basic parameters, including position, pulsation period, mean \textit{I}-band brightness, and \textit{I}-band amplitude. Phase-folded \textit{I}-band light curves for the classified objects are presented in this paper, while the full time-series data and finding charts are available in the online version of the OGLE Internet Archive:
\begin{center}
\textit{https://www.astrouw.edu.pl/ogle/ogle4/OCVS/}
\end{center}
\begin{center}
\textit{https://ogledb.astrouw.edu.pl/~ogle/OCVS/}
\end{center}
For the most recent additions to the sample, we adopt the same naming scheme as in the previous OGLE catalogs and BLAP studies: BLAPs in the inner Galactic bulge are numbered from 002 to 037, those in the Galactic disk are 001 and from 038 to 061, and those in the outer Galactic bulge from 062 to 094. The newly discovered objects are labeled sequentially as OGLE-BLAP-NNN, starting with NNN = 095 and ordered by increasing right ascension up to OGLE-BLAP-184.

Observational parameters of the newly found BLAPs are listed in Table~1. In the table, we provide the variable name, J2000 equatorial coordinates, galactic coordinates, pulsation period, mean \textit{I}-band brightness, $(V-I)$ color index, \textit{I}-band amplitude, and \textit{V}-band amplitude in cases where it could be determined. For some stars, the $(V-I)$ colors are unavailable because no \textit{V}-band measurements exist in the OGLE database. This mainly reflects the fact that some objects are fainter than the survey depth in \textit{V}, especially in heavily reddened fields. In the case of objects which show period variations during the timespan of observations, we provide mean period calculated on the full available dataset for a given star. 

Fig.~1 shows a map with the locations of all BLAPs known to date. Objects that are part of OGLE collection are marked with blue dots, while objects that were not detected/classified by OGLE are marked in green. Panel A in Fig.~1 provides a zoom-in of the Galactic bulge area, whereas panel B presents a map on the scale of the entire Galactic plane. The OGLE-IV inner BLG fields analyzed in this study are highlighted in red. Fields of previous searches in the OGLE data are shown in white (Galactic disk fields, Borowicz \etal 2023a) and grey (outer Galactic bulge fields Borowicz \etal 2023b).

The map clearly shows that BLAPs are concentrated both along the Galactic plane and around the Galactic center. However, this should be interpreted with caution, as the completeness of the searches is likely affected by a number of data points, since OGLE monitors the bulge deeper and more frequently than the disk. More detailed analysis on how the number of observations affects the completeness of the catalog is presented in Section~7. The majority of the pulsators are detected within a $\pm 10^\circ$ strip along the Galactic plane. The lack of BLAPs within  $- 2^\circ < l < + 2^\circ$ is caused by extinction, pushing sources below the detection limit. The same effect can be noted in catalogs of other types of variables; for example, RR Lyrae stars (Soszy\'nski \etal 2019). 
\begin{figure}
\centerline{\includegraphics[width=15cm]{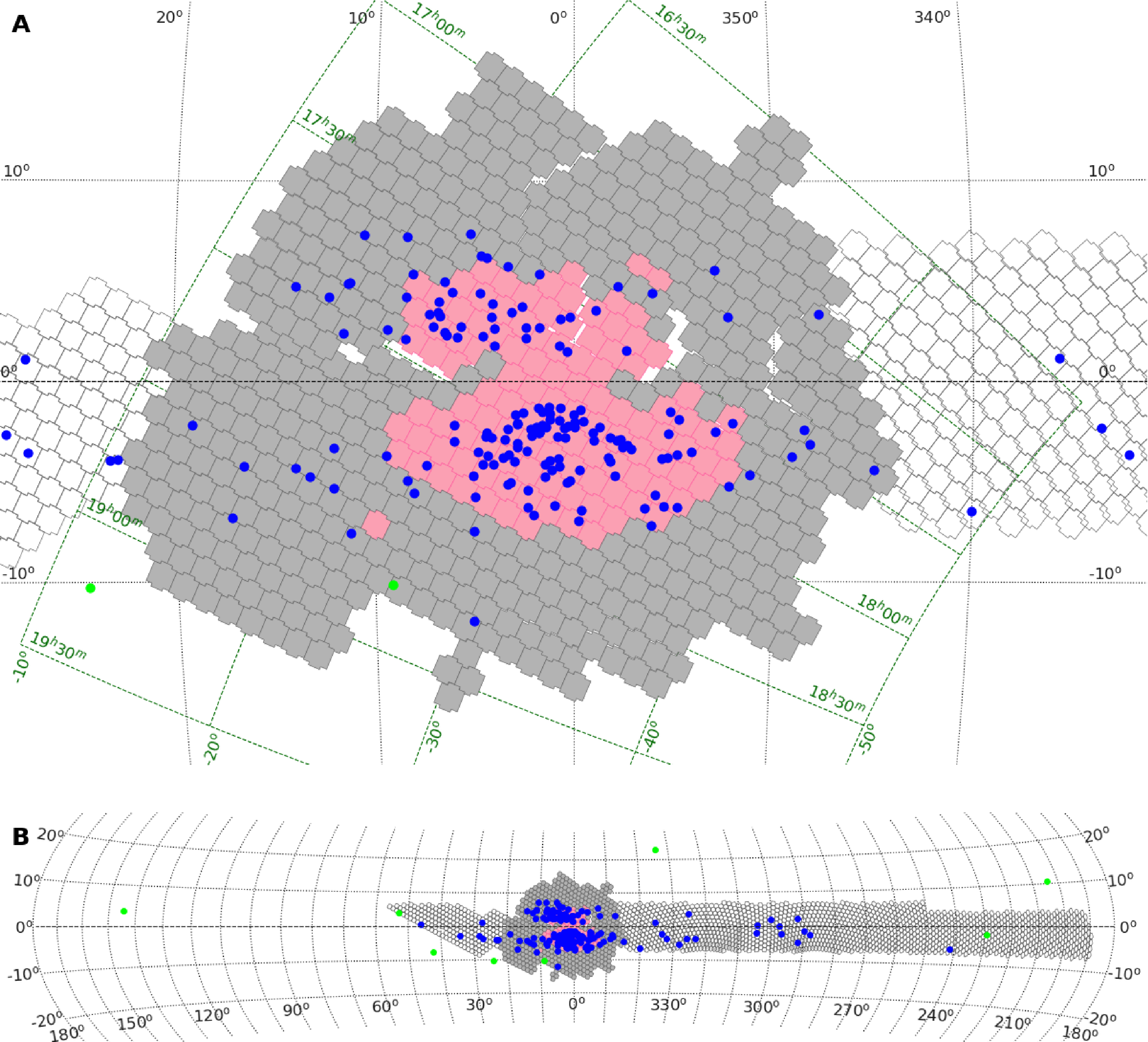}}
\FigCap{On-sky distribution of all known BLAPs. Objects that are part of the OGLE collection are marked with blue dots. Green dots denote positions of BLAPs that do not have OGLE counterparts. The fields investigated in previous OGLE searches are displayed in white for the Galactic disk (Borowicz \etal 2023a) and in grey for the outer Galactic bulge (Borowicz \etal 2023b). Fields that are examined in this study are highlighted in red.}
\end{figure}

\begin{table}[]
\scriptsize
\caption{List of newly detected BLAPs in the OGLE inner bulge fields}
\begin{tabular}{@{}lccrrrccccc@{}}
\toprule
\multicolumn{1}{c}{Name}            & RA (J2000)         & Dec (J2000)      & \multicolumn{1}{c}{$l$}        & \multicolumn{1}{c}{$b$}        & Period    & \textit{I}       & $(V-I)$       & $A_I$      & $A_V$   \\             
              &      {[$^\circ$]}      &  {[$^\circ$]}         &   \multicolumn{1}{c}{[$^\circ$]}         &   \multicolumn{1}{c}{[$^\circ$]}         & {[}min{]} & {[mag]}  & {[}mag{]}   & {[}mag{]}  & {[}mag{]}  \\ 
\midrule
OGLE-BLAP-095 & 259.72975  &$-$29.80631 &$-$3.90745 & 4.39868  & 18.60 & 18.913  & 0.922 & 0.222 & 0.198 \\ 
OGLE-BLAP-096 & 260.51313  &$-$28.21564 &$-$2.20352 & 4.73713  & 40.58 & 18.438  & 1.099 & 0.216 &$-$     \\
OGLE-BLAP-097 & 262.32418  &$-$27.98440 &$-$1.11887 & 3.53906  & 31.19 & 18.346  & 1.451 & 0.217 & 0.270 \\
OGLE-BLAP-098 & 263.15193  &$-$21.97065 & 4.35385  & 6.18431  & 57.15 & 15.618  & 1.415 & 0.060 &$-$     \\
OGLE-BLAP-099 & 263.25681  &$-$21.66428 & 4.66635  & 6.26697  & 18.13 & 18.824  & 0.921 & 0.216 &$-$     \\
OGLE-BLAP-100 & 263.27095  &$-$30.34458 &$-$2.64298 & 1.55521  & 61.34 & 17.727  & 1.662 & 0.146 &$-$     \\
OGLE-BLAP-101 & 263.41532  &$-$27.07275 & 0.17543  & 3.22610  & 27.44 & 19.604  & 2.297 & 0.260 &$-$     \\
OGLE-BLAP-102 & 263.47047  &$-$36.76823 &$-$7.94392 &$-$2.07809 & 33.06 & 19.604  & 1.789 & 0.297 &$-$     \\
OGLE-BLAP-103 & 263.81667  &$-$26.72812 & 0.65953  & 3.11105  & 39.18 & 18.538  & 1.388 & 0.316 & 0.368 \\
OGLE-BLAP-104 & 264.39266  &$-$24.77683 & 2.58835  & 3.71894  & 20.14 & 18.548  & 1.334 & 0.213 &$-$     \\
OGLE-BLAP-105 & 264.54677  &$-$36.26696 &$-$7.05529 &$-$2.53752 & 30.61 & 18.645  & 1.290 & 0.176 &$-$     \\
OGLE-BLAP-106 & 264.86760  &$-$26.08187 & 1.71033  & 2.66130  & 24.33 & 19.030  & 1.637 & 0.216 &$-$     \\
OGLE-BLAP-107 & 264.94119  &$-$33.83215 &$-$4.82108 &$-$1.51420 & 67.95 & 17.788  & 2.007 & 0.140 &$-$     \\
OGLE-BLAP-108 & 264.94773  &$-$24.49953 & 3.09232  & 3.43885  & 26.13 & 18.235  & 1.320 & 0.128 &$-$     \\
OGLE-BLAP-109 & 265.07544  &$-$34.41713 &$-$5.25807 &$-$1.91887 & 38.56 & 18.787  & 1.752 & 0.198 &$-$     \\
OGLE-BLAP-110 & 265.08881  &$-$23.44543 & 4.05728  & 3.88652  & 32.39 & 17.689  & 0.850 & 0.091 &$-$     \\
OGLE-BLAP-111 & 265.11927  &$-$27.40810 & 0.70453  & 1.76728  & 40.62 & 19.240 &$-$     & 0.199 &$-$     \\
OGLE-BLAP-112 & 265.14955  &$-$27.88014 & 0.31825  & 1.49441  & 33.54 & 19.220 &$-$     & 0.230 &$-$     \\
OGLE-BLAP-113 & 265.25553  &$-$25.50481 & 2.38553  & 2.67056  & 28.54 & 18.296  & 1.410 & 0.309 &$-$     \\
OGLE-BLAP-114 & 265.69551  &$-$21.42304 & 6.07844  & 4.47058  & 24.14 & 19.500 &$-$     & 0.379 &$-$     \\
OGLE-BLAP-115 & 265.74576  &$-$23.76906 & 4.09848  & 3.20431  & 54.58 & 17.397  & 1.150 & 0.163 &$-$     \\
OGLE-BLAP-116 & 265.75147  &$-$25.76127 & 2.40288  & 2.15539  & 14.85 & 19.988  &$-$     & 0.381 &$-$     \\
OGLE-BLAP-117 & 266.10606  &$-$34.33477 &$-$4.74025 &$-$2.59903 & 18.73 & 18.802  & 0.929 & 0.365 & 0.357 \\
OGLE-BLAP-118 & 266.32102  &$-$35.79557 &$-$5.89774 &$-$3.50949 & 72.28 & 17.037  & 0.677 & 0.109 & 0.120 \\
OGLE-BLAP-119 & 266.52287  &$-$21.12790 & 6.73286  & 3.96505  & 48.07 & 18.385  & 1.668 & 0.120 &$-$     \\
OGLE-BLAP-120 & 266.88946  &$-$23.91023 & 4.52246  & 2.23739  & 17.49 & 18.641  & 1.714 & 0.323 &$-$     \\
OGLE-BLAP-121 & 266.97494  &$-$38.27466 &$-$7.75926 &$-$5.23385 & 12.62 & 18.830  & 0.386 & 0.396 &$-$     \\
OGLE-BLAP-122 & 267.00905  &$-$24.64443 & 3.94961  & 1.76497  & 32.26 & 19.378 &$-$     & 0.266 &$-$     \\
OGLE-BLAP-123 & 267.07117  &$-$22.68357 & 5.66062  & 2.72685  & 19.02 & 18.731  & 1.259 & 0.219 &$-$     \\
OGLE-BLAP-124 & 267.13741  &$-$21.53147 & 6.68176  & 3.26651  & 21.41 & 18.371  & 1.061 & 0.183 &$-$     \\
OGLE-BLAP-125 & 267.36832  &$-$21.00662 & 7.24365  & 3.35081  & 18.08 & 18.127  & 1.087 & 0.212 &$-$     \\
OGLE-BLAP-126 & 267.59170  &$-$29.93420 &$-$0.32052 &$-$1.40051 & 24.45 & 19.510 &$-$     & 0.173 &$-$     \\
OGLE-BLAP-127 & 267.62961  &$-$22.81532 & 5.81187  & 2.21719  & 21.49 & 18.419  & 1.158 & 0.135 &$-$     \\
OGLE-BLAP-128 & 267.63384  &$-$34.71743 &$-$4.41886 &$-$3.87485 & 10.41 & 18.428  & 0.562 & 0.106 & 0.133 \\
OGLE-BLAP-129 & 267.84809  &$-$21.49968 & 7.04770  & 2.71460  & 34.81 & 18.522  & 1.568 & 0.323 &$-$     \\
OGLE-BLAP-130 & 267.85532  &$-$22.30661 & 6.35613  & 2.29768  & 32.00 & 19.273  & 1.852 & 0.425 &$-$     \\
OGLE-BLAP-131 & 267.87160  &$-$32.42473 &$-$2.34098 &$-$2.87825 & 19.36 & 18.446  & 2.211 & 0.139 & 0.223 \\
OGLE-BLAP-132 & 267.91870  &$-$31.20301 &$-$1.26799 &$-$2.29067 & 24.24 & 18.964  & 2.061 & 0.188 &$-$     \\
OGLE-BLAP-133 & 268.00696  &$-$29.79084 &$-$0.01322 &$-$1.63690 & 22.90 & 19.719 &$-$     & 0.431 &$-$     \\
OGLE-BLAP-134 & 268.01268  &$-$32.85740 &$-$2.65366 &$-$3.20047 & 9.45  & 19.890  & 1.996 & 0.132 &$-$     \\
OGLE-BLAP-135 & 268.06548  &$-$33.13540 &$-$2.87114 &$-$3.37965 & 14.50 & 19.656  & 0.680 & 0.180 &$-$     \\
OGLE-BLAP-136 & 268.06989  &$-$32.29916 &$-$2.14746 &$-$2.95868 & 27.14 & 18.624  & 1.618 & 0.224 &$-$     \\
OGLE-BLAP-137 & 268.08817  &$-$30.38331 &$-$0.48773 &$-$1.99887 & 12.36 & 19.253 &$-$     & 0.139 &$-$     \\
OGLE-BLAP-138 & 268.10251  &$-$29.02166 & 0.69144  &$-$1.31698 & 27.53 & 19.689 &$-$     & 0.324 &$-$     \\
OGLE-BLAP-139 & 268.22288  &$-$32.70611 &$-$2.43333 &$-$3.27614 & 24.36 & 18.498  & 0.898 & 0.159 & 0.148 \\
OGLE-BLAP-140 & 268.36748  &$-$28.52579 & 1.23669  &$-$1.26515 & 13.66 & 19.118  & 2.854 & 0.146 &$-$     \\
OGLE-BLAP-141 & 268.37423  &$-$31.07806 &$-$0.96215 &$-$2.56316 & 36.31 & 18.867  & 1.274 & 0.198 & 0.166 \\
OGLE-BLAP-142 & 268.63295  &$-$30.12901 &$-$0.02981 &$-$2.27581 & 6.72  & 18.968  & 1.295 & 0.153 &$-$     \\
OGLE-BLAP-143 & 268.71397  &$-$28.72817 & 1.21595  &$-$1.62987 & 37.24 & 17.362  & 1.938 & 0.099 & 0.192 \\
OGLE-BLAP-144 & 268.71502 &$-$31.31832 &$-$1.02255 &$-$2.93608 & 18.74 & 18.946  & 1.392 & 0.095 &$-$     \\
OGLE-BLAP-145 & 268.75465  &$-$28.11013 & 1.76779  &$-$1.34892 & 23.01 & 19.841 &$-$     & 0.290 &$-$     \\
OGLE-BLAP-146 & 268.76920  &$-$29.31408 & 0.73412  &$-$1.96705 & 43.45 & 17.822  & 1.234 & 0.202 &$-$     \\
OGLE-BLAP-147 & 268.79328  &$-$28.36440 & 1.56533  &$-$1.50655 & 29.79 & 19.333 &$-$     & 0.370 &$-$     \\
OGLE-BLAP-148 & 268.99175 &$-$29.70985 & 0.48933  &$-$2.33353 & 7.40  & 19.498  & 0.903 & 0.131 &$-$     \\
OGLE-BLAP-149 & 269.30706  &$-$32.53896 &$-$1.82924 &$-$3.98205 & 18.99 & 18.327 &$-$     & 0.128 &$-$     \\
OGLE-BLAP-150 & 269.42381  &$-$27.56825 & 2.53382  &$-$1.58837 & 18.57 & 19.231  & 2.288 & 0.342 &$-$     \\
\bottomrule
\end{tabular}
\end{table}
\begin{table}[]
\scriptsize
\caption{List of newly detected BLAPs in the OGLE inner bulge fields, continued}
\begin{tabular}{@{}lccrrrcccccc@{}}
\toprule
\multicolumn{1}{c}{Name}            & RA (J2000)       & Dec (J2000)       & \multicolumn{1}{c}{$l$}        & \multicolumn{1}{c}{$b$}        & Period    & \textit{I}       & $(V-I)$       & $A_I$      & $A_V$   \\             
              & {[$^\circ$]}     & {[$^\circ$]}    & \multicolumn{1}{c}{[$^\circ$]}     & \multicolumn{1}{c}{[$^\circ$]}     & {[}min{]} & {[mag]}   & {[}mag{]}   & {[}mag{]}  & {[}mag{]} \\ \midrule
OGLE-BLAP-151 & 269.43187  &$-$28.90193 & 1.38156  &$-$2.26110 & 34.62 & 17.923  & 0.871 & 0.273 &$-$     \\
OGLE-BLAP-152 & 269.43757  &$-$29.99040 & 0.43993  &$-$2.80870 & 17.95 & 18.483  & 0.962 & 0.094 & 0.123 \\
OGLE-BLAP-153 & 269.49647  &$-$28.68342 & 1.59931  &$-$2.20105 & 5.36  & 18.647  & 1.318 & 0.108 &$-$     \\
OGLE-BLAP-154 & 269.88857  &$-$28.76622 & 1.69873  &$-$2.54050 & 20.76 & 18.063  & 0.910 & 0.126 & 0.137 \\
OGLE-BLAP-155 & 269.96606  &$-$28.70147 & 1.78875  &$-$2.56737 & 15.00 & 18.147  & 0.715 & 0.176 &$-$     \\
OGLE-BLAP-156 & 270.00213  &$-$28.29683 & 2.15621  &$-$2.39427 & 31.79 & 17.812  & 0.884 & 0.235 &$-$     \\
OGLE-BLAP-157 & 270.31479 &$-$28.48771 & 2.12635  &$-$2.72774 & 7.87  & 18.808  & 0.449 & 0.094 &$-$     \\
OGLE-BLAP-158 & 270.58616  &$-$27.30661 & 3.27332  &$-$2.35417 & 20.32 & 18.376  & 1.880 & 0.209 & 0.204 \\
OGLE-BLAP-159 & 270.68986  &$-$30.20766 & 0.78706  &$-$3.85795 & 10.54 & 18.529 &$-$     & 0.241 &$-$     \\
OGLE-BLAP-160 & 270.78887  &$-$35.17922 &$-$3.53351 &$-$6.34930 & 26.31 & 16.967  & 0.111 & 0.192 &$-$     \\
OGLE-BLAP-161 & 271.03512  &$-$29.88030 & 1.21970  &$-$3.95883 & 60.18 & 16.213  & 0.398 & 0.114 & 0.101 \\
OGLE-BLAP-162 & 271.05704  &$-$30.45655 & 0.72437  &$-$4.25604 & 9.31  & 18.184  & 1.015 & 0.099 & 0.129 \\
OGLE-BLAP-163 & 271.17789  &$-$28.65310 & 2.35435  &$-$3.46941 & 11.21 & 17.418  & 0.944 & 0.072 &$-$     \\
OGLE-BLAP-164 & 271.22943  &$-$27.44662 & 3.43146  &$-$2.92069 & 14.25 & 17.853  & 1.106 & 0.081 & 0.107 \\
OGLE-BLAP-165 & 271.31129  &$-$29.78292 & 1.42183  &$-$4.12078 & 9.29  & 19.250  & 0.264 & 0.245 & 0.272 \\
OGLE-BLAP-166 & 271.37799  &$-$26.45378 & 4.36382  &$-$2.55196 & 32.76 & 17.491  & 0.833 & 0.269 &$-$     \\
OGLE-BLAP-167 & 271.46169  &$-$31.27995 & 0.17078  &$-$4.95898 & 9.90  & 18.072  & 0.768 & 0.132 & 0.180 \\
OGLE-BLAP-168 & 271.67137  &$-$31.18515 & 0.34092  &$-$5.07036 & 13.47 & 18.406  & 0.612 & 0.276 &$-$     \\
OGLE-BLAP-169 & 271.89280  &$-$24.85511 & 5.98771  &$-$2.17973 & 20.51 & 19.273  & 2.035 & 0.291 &$-$     \\
OGLE-BLAP-170 & 272.01894  &$-$28.38216 & 2.95077  &$-$3.98450 & 11.27 & 19.395  & 0.744 & 0.439 &$-$     \\
OGLE-BLAP-171 & 272.17697  &$-$27.76494 & 3.56024  &$-$3.80942 & 9.02  & 18.454  & 1.034 & 0.187 & 0.288 \\
OGLE-BLAP-172 & 272.47433  &$-$27.03931 & 4.32507  &$-$3.69207 & 13.14 & 19.081  & 0.961 & 0.126 & 0.125 \\
OGLE-BLAP-173 & 272.52611  &$-$26.56142 & 4.76735  &$-$3.50317 & 19.27 & 18.609  & 0.818 & 0.214 & 0.169 \\
OGLE-BLAP-174$^{(1)}$ & 272.75092  &$-$27.50369 & 4.03449  &$-$4.13030 & 22.76 & 17.561  & 0.403 & 0.092 &$-$     \\
OGLE-BLAP-175 & 273.02671  &$-$27.06479 & 4.53847  &$-$4.13629 & 38.05 & 16.973  & 0.546 & 0.262 &$-$     \\
OGLE-BLAP-176$^{(2)}$ & 273.11615  &$-$29.64677 & 2.29552  &$-$5.43136 & 10.85 & 18.127  & 0.318 & 0.268 &$-$     \\
OGLE-BLAP-177 & 273.18257  &$-$28.72208 & 3.14107  &$-$5.04481 & 6.39  & 18.222  & 0.561 & 0.133 &$-$     \\
OGLE-BLAP-178 & 273.33703  &$-$32.58519 &$-$0.22402 &$-$6.97943 & 9.79  & 18.196  & 0.229 & 0.206 &$-$     \\
OGLE-BLAP-179 & 273.34956  &$-$28.62111 & 3.29997  &$-$5.12612 & 6.13  & 18.912  & 0.110 & 0.252 &$-$     \\
OGLE-BLAP-180 & 273.84972  &$-$26.92113 & 5.01390  &$-$4.71401 & 15.57 & 17.998  & 0.412 & 0.252 &$-$     \\
OGLE-BLAP-181 & 274.01630  &$-$30.06103 & 2.29628  &$-$6.31515 & 76.36 & 14.478  & 0.089 & 0.096 &$-$     \\
OGLE-BLAP-182 & 274.23339  &$-$30.44902 & 2.03831  &$-$6.66156 & 9.46  & 18.677  & 0.164 & 0.225 &$-$     \\
OGLE-BLAP-183 & 274.54977  &$-$24.59110 & 7.37375  &$-$4.17451 & 21.52 & 17.853  & 0.572 & 0.165 &$-$     \\
OGLE-BLAP-184 & 275.10614  &$-$22.61254 & 9.36515  &$-$3.69743 & 9.29  & 18.396  & 0.727 & 0.169 &$-$     \\ \bottomrule 
\end{tabular}
\vspace{0.25em} 

\scriptsize{$(1)$ =OW-BLAP-2 (Ramsay \etal 2022), $(2)$ =OW-BLAP-1 (Ramsay \etal 2022)}
\end{table}
\begin{figure}
\centerline{\includegraphics[width=13.7cm]{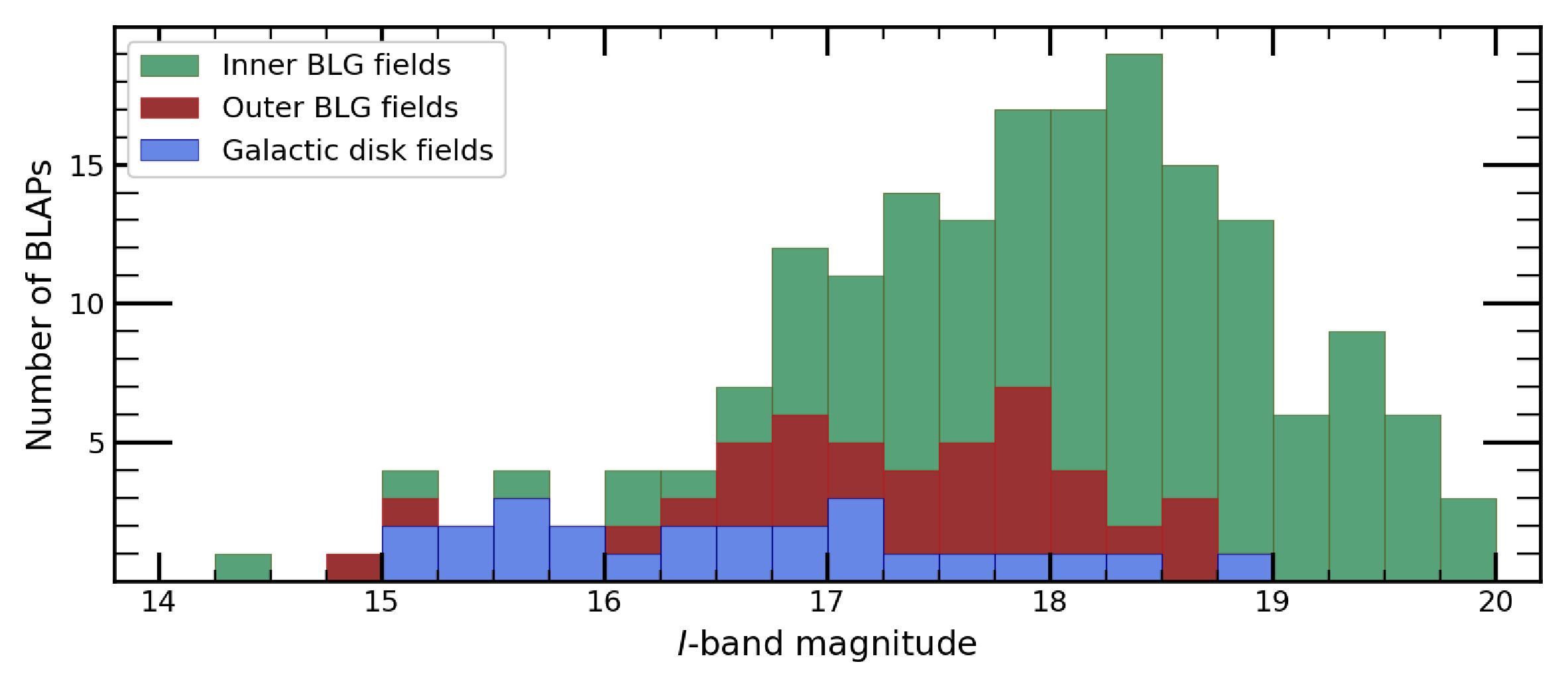}}
\FigCap{Distribution of \textit{I}-band magnitudes for all 184 BLAPs detected in the OGLE data.}
\end{figure}

Fig.~2 presents the distribution of mean \textit{I}-band magnitudes for all BLAPs observed by OGLE, grouped according to the survey fields. BLAPs identified in the Galactic disk fields (Borowicz \etal 2023a) are shown in blue, while those from the outer bulge fields (Borowicz \etal 2023b) are marked in red. Thirteen original pulsators reported in Pietrukowicz \etal (2017) (excluding OGLE-BLAP-001, which resides in the disk), together with additional objects located in the inner bulge fields from Pietrukowicz \etal (2025a), and those analyzed in this work, are highlighted in green. In the case of the inner bulge fields, based on the distribution shown in Fig.~2, we conclude that the survey is highly complete down to $I=18.5$ mag. As part of the shallow GVS survey, the Galactic disk and outer bulge fields were observed with shorter exposures, the majority of which were obtained through the \textit{I}-band filter with an integration time of 25 s. The survey depth in these regions is shallower than in the inner BLG fields by about 1~mag. 
\begin{figure}
\centerline{\includegraphics[width=13.7cm]{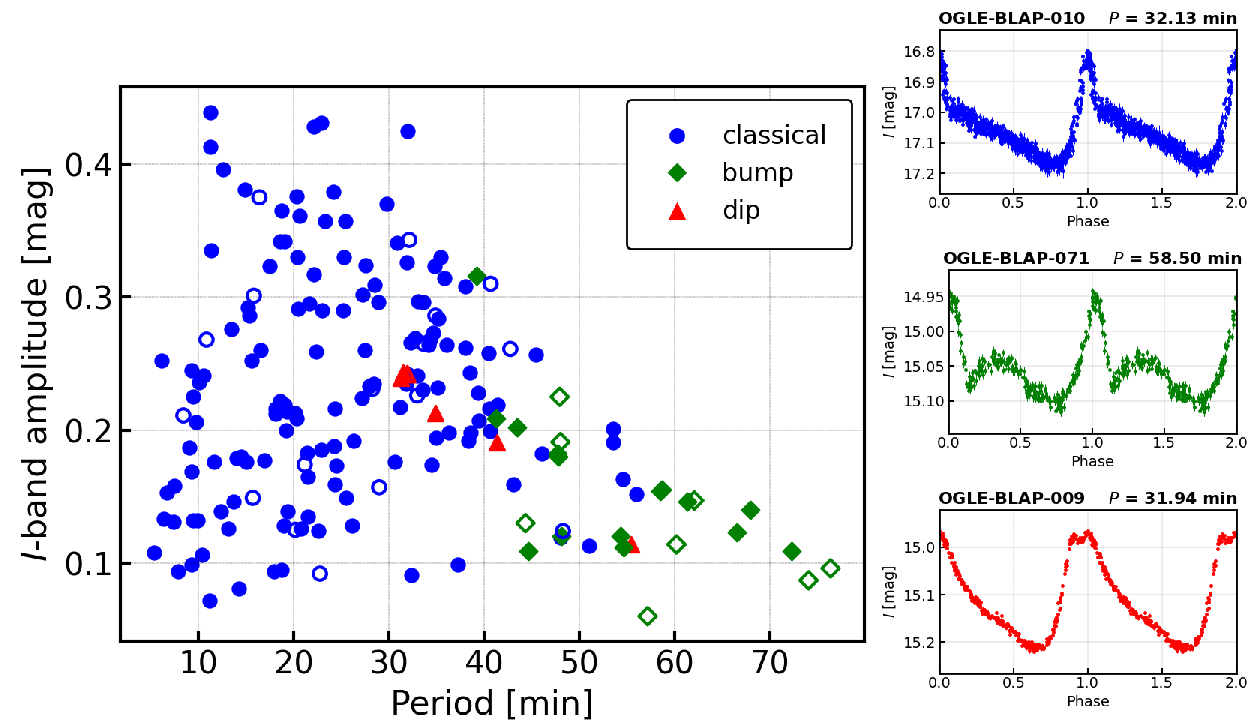}}
\FigCap{\textit{I}-band amplitude vs. period diagram for all BLAPs detected by OGLE. Blue circles mark BLAPs with "classical" light curve shape. Red triangles indicate objects that exhibit a dimming near the maximum brightness. BLAPs with an additional bump in the light curve are marked with green diamonds. Objects with available spectroscopic data are represented with open symbols. Representative light curve examples are shown in the right panels.}
\end{figure}
Fig.~3 presents the period-amplitude diagram for all 184 BLAPs observed by OGLE. The peak-to-peak amplitudes were determined by finding the best Fourier fits to the phased light curves. We distinguish here three principal types of light curves. Stars exhibiting the classical sawtooth morphology, characteristic of radially pulsating variables, are marked in blue. BLAPs whose light curves display an additional bump covering roughly half of the pulsation cycle are shown in green. Examples include objects OGLE-BLAP-098, OGLE-BLAP-146, and OGLE-BLAP-181. In red, we highlight pulsators whose light curves reveal a dip near the maximum brightness, with OGLE-BLAP-009 (Pietrukowicz \etal 2017, Bradhsaw \etal 2024) being the best-known representative of this group. We note that, for some of the newly discovered BLAPs, their low brightness near the survey detection limit may have prevented the identification of subtle light-curve features, such as shallow dips. The \textit{I}-band amplitudes of the newly identified BLAPs span the range from about 0.06 to 0.44 mag. One has to remember that light curve shapes of variable stars at the short-period end are distorted and their amplitudes are reduced due to the fixed exposure time. Fig.~3 also demonstrates that additional light-curve features occur predominantly toward the long-period end, with the bumps becoming apparent at pulsation periods above approximately 40 min. Objects confirmed spectroscopically, either in this or in previous studies, are indicated with open symbols. We also note that some of the classified BLAPs may be affected by blending, which results in a reduction of the observed amplitude. A clear example is OGLE-BLAP-143 ($P = 37.24$ min, $A_I = 0.099$ mag). The light curve of this star exhibits a characteristic morphology of BLAPs, but the color is unusually red and the amplitude is likely underestimated.
\begin{figure}
\centerline{\includegraphics[width=13.7cm]{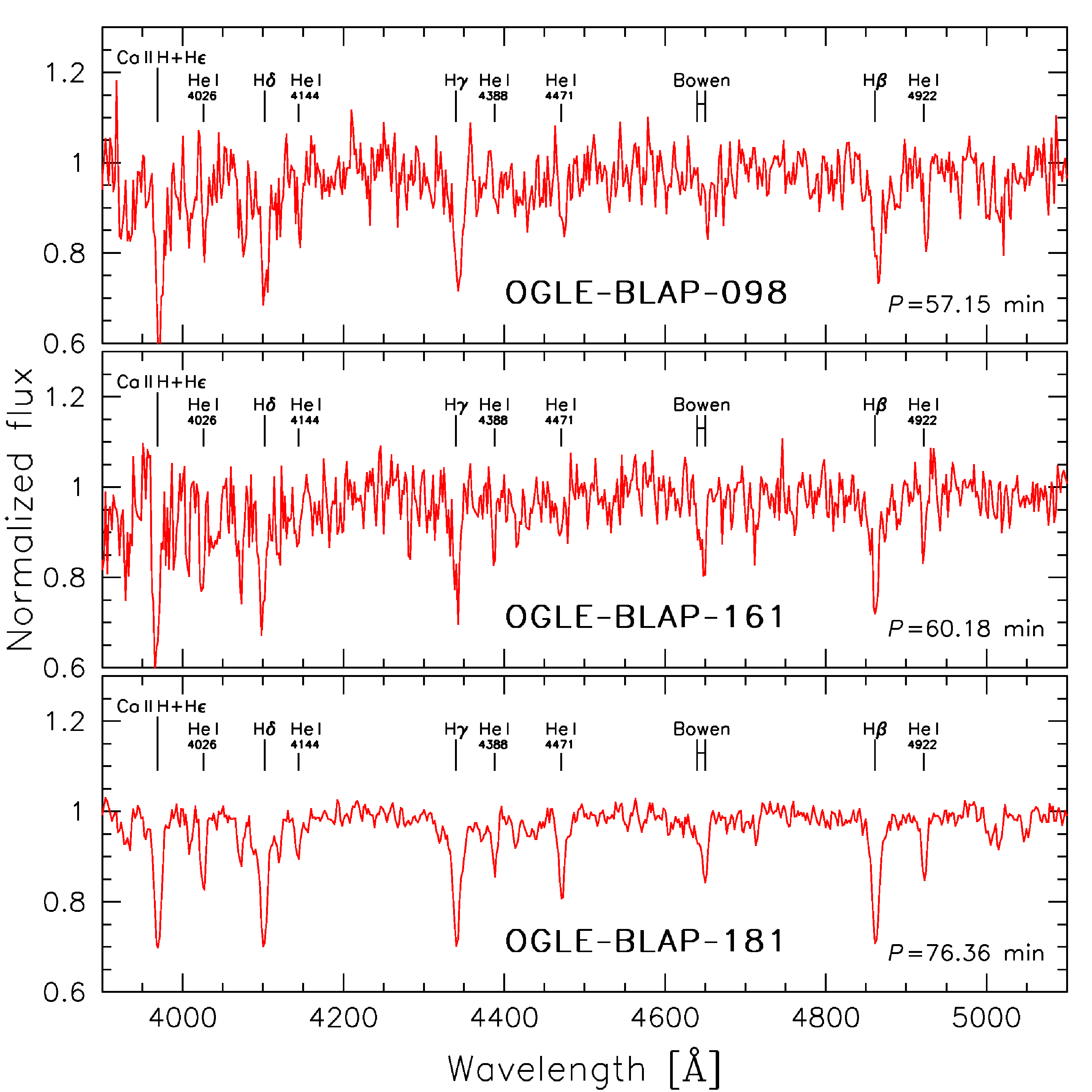}}
\FigCap{Low-resolution spectra of three newly identified BLAPs, obtained with EFOSC2 at NTT. }
\end{figure}
For three of the newly discovered BLAPs that displayed unusual properties, we obtained low-resolution spectra in order to verify their classification. The final normalized spectra, produced by co-adding two single consecutive exposures, are presented in Fig.~4. The first of these objects, OGLE-BLAP-098, although its light curve resembles that of a typical long-period BLAP (e.g., OGLE-BLAP-071), exhibits an unusually low \textit{I}-band amplitude of 0.06~mag and a substantial scatter in the phase-folded light curve, which cannot be easily attributed to either monotonic or cyclic period changes. The remaining two objects, OGLE-BLAP-161 and OGLE-BLAP-181, show strongly asymmetric light curves characterized by sharp minima reminiscent of eclipses. Their light-curve morphology is clearly distinct from the typical bump observed in long-period BLAPs. The spectra of all three stars show features characteristic of helium-rich BLAPs, with prominent He~I lines at 4471\,\AA\ and 4922\,\AA. Based on these spectra, the three objects were classified as BLAPs despite initial concerns regarding the nature of these variable objects. However, given the low resolution of the spectra, we did not attempt to determine physical parameters, as any such estimates would be highly uncertain.
\begin{figure}
\centerline{\includegraphics[width=13.7cm]{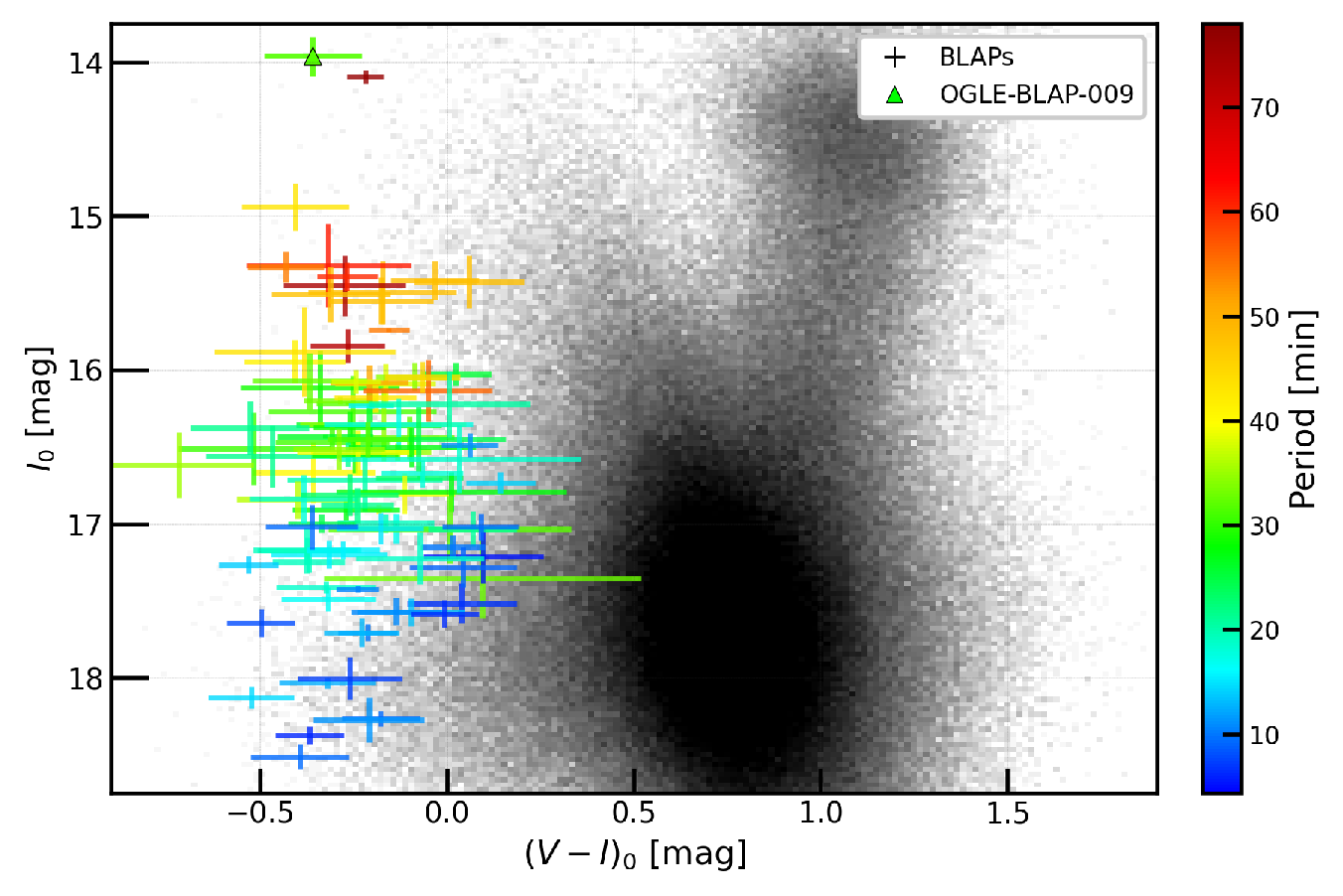}}
\FigCap{Dereddened $I_0$ vs. $(V-I)_{0}$ color-magnitude diagram of BLAPs from the inner Galactic bulge fields. Dereddening was performed using OGLE-III extinction maps (Nataf \etal 2013) and OGLE RR Lyrae stars (Pietrukowicz \etal 2015). Colors correspond to the measured period. OGLE-BLAP-009, a foreground (Galactic disk) object, is highlighted. The background stars were selected from the inner BLG fields of the OGLE-IV survey.}
\end{figure}

Previous studies on the spatial distribution of BLAPs suggested that these stars are distant objects located several kiloparsecs from the Sun (Pietrukowicz \etal 2025a), and most of them are associated with the Galactic bulge. Several lines of evidence support this view: the on-sky distribution, very small parallaxes near the measurement limits, the proper motions of the stars, and the previously determined luminosities, for example that of OGLE-BLAP-009 (Bradshaw \etal 2024; Pietrukowicz \etal 2025a). Since the sample in the above study originates from dense bulge fields and includes relatively faint detections, it was of limited utility for Gaia parallax measurements. By applying a distance-independent dereddening procedure, it was possible to place all OGLE BLAPs on a reddening-free color-magnitude 
diagram, shown in Fig.~5. Two dereddening methods were used. If the star's position was covered by the OGLE-III, we applied reddening maps (Nataf \etal 2013) constructed based on red clump stars from the bulge, providing the color excess $E(V-I)$ and absorption value $A_I$. For BLAPs located outside the OGLE-III coverage, the extinction parameters were estimated using RR Lyrae stars from the OGLE-IV collection (Soszy{\'n}ski \etal 2019), relying on the period-luminosity relation (Pietrukowicz \etal 2015). The parameters were derived for all RR Lyrae stars within a 5 arcmin radius observed around a target BLAP and then a distance weighted average was determined. We tried applying the above dereddening procedures to the whole catalog of 184 OGLE BLAPs. However, the dereddened values of $I_0$ and $(V-I)_0$ could be derived only for around 100 stars, since either no nearby RR Lyrae stars were available or the corresponding regions were not covered by the OGLE-III maps. Variables that are suspected to be affected by blending, like OGLE-BLAP-143, are not included in Fig.~5. For OGLE-BLAP-009, which is located in the disk, dereddened brightness and color index were taken from Pietrukowicz \etal (2025a), as the above procedures would result in an overcorrection. In Fig.~5, the objects are coded with colors corresponding to their pulsation periods. The main sources of uncertainty in the position of the stars in this diagram are the uncertainty in determining the \textit{V}-band brightness and the uncertainties associated with estimating the extinction parameters. Additionally, there is an error related to the distance-independent extinction determination, although this is difficult to quantify. Since BLAPs follow a period-gravity relation, they are also expected to obey a period-luminosity relation (Pietrukowicz \etal 2025a). Taking this into account, the appearance of Fig.~5 is not surprising; objects with longer periods are generally brighter, which is clearly visible in the figure. This further reinforces the conclusion that they belong to the Galactic bulge.

Among the newly identified variable objects, two stars were previously reported in the literature as BLAPs, namely OGLE-BLAP-174 as OW-BLAP-2 and OGLE-BLAP-176 as OW-BLAP-1 (Ramsay \etal 2022). A prior independent classification of OGLE-BLAP-156 based on multiband photometry and OGLE-III data was reported by Bru\'{s} (2022). The remaining stars are introduced here for the first time as BLAPs. The present study further extends the known range of BLAP pulsation periods. A new record holder for the shortest-period OGLE BLAP is OGLE-BLAP-153 with $P = 5.36$ min. However, we should note that even shorter-period BLAPs are known, \eg HG-BLAP-1 (3.33 min, Kupfer \etal 2019). At the opposite end of the period distribution is OGLE-BLAP-181 ($P = 76.36$ min) which replaces OGLE-BLAP-022 ($P = 74.05$ min) as the longest-period BLAP known so far. The newly detected BLAPs significantly increase the sample size at both the short- and long-period ends of the distribution: 14 BLAPs exhibit periods shorter than 10~min, while five BLAPs have periods longer than 60~min. Additionally, several objects were found with intermediate periods bridging the range between original BLAPs and high-gravity BLAPs, further confirming that the previously claimed period gap is not an intrinsic property but rather a consequence of the small number of objects known at the time. We note that variable OGLE-BLAP-149 was classified by Gaia as an eclipsing binary candidate (source\_id: 4043360350881337216, Mowlavi \etal 2023), but the provided frequency ($f = 4.046$ c/d) is not present in the OGLE data.

Phase-folded \textit{I}-band light curves of all 90 BLAPs reported in this work are presented in Figs.~6--10. For clarity, the light curves were derived from the first 2000 observations, or from all available data when fewer measurements were present. Photometric uncertainties are indicated in the plots. The OGLE-IV dataset covers the years 2010--2024, while for six stars, only OGLE-III observations from 2001--2009 are available.

Among the stars analyzed in this study, we identify several distinct types of light-curve morphology. In addition to the classical sawtooth shape (e.g., OGLE-BLAP-113, OGLE-BLAP-129, OGLE-BLAP-151), a number of objects display an additional bump, a feature typically associated with longer-period BLAPs, as seen in OGLE-BLAP-100 and OGLE-BLAP-107, for instance. Some stars exhibit an "inverted" shape, where the rise to the maximum brightness is slower than the decline, with OGLE-BLAP-118 as an example. A similar behavior was also observed in the previously identified OGLE-BLAP-029 and OGLE-BLAP-069. Another characteristic pattern involves a very narrow maximum, followed by nearly constant brightness throughout the remaining phase, with OGLE-BLAP-115 providing the clearest example. Its morphology and period closely resemble those of ZGP-BLAP-10. We also note striking similarities between OGLE-BLAP-096 ($P = 40.58$ min) and OGLE-BLAP-031 ($P = 40.67$ min), both in period and light-curve shape. Finally, we point out that OGLE-BLAP-174 (OW-BLAP-2), similarly to OGLE-BLAP-093, while spectroscopically confirmed BLAPs, exhibits light curves that cannot be consistently phased with their derived periods. Their behavior does not appear to be readily explained by multimodality or a constant rate of period change. For some of the faintest BLAPs in our sample, only limited structure can be recognized in their light curves, as these stars have brightness close to the survey detection limit.
\begin{figure}
\centerline{\includegraphics[width=13.71cm]{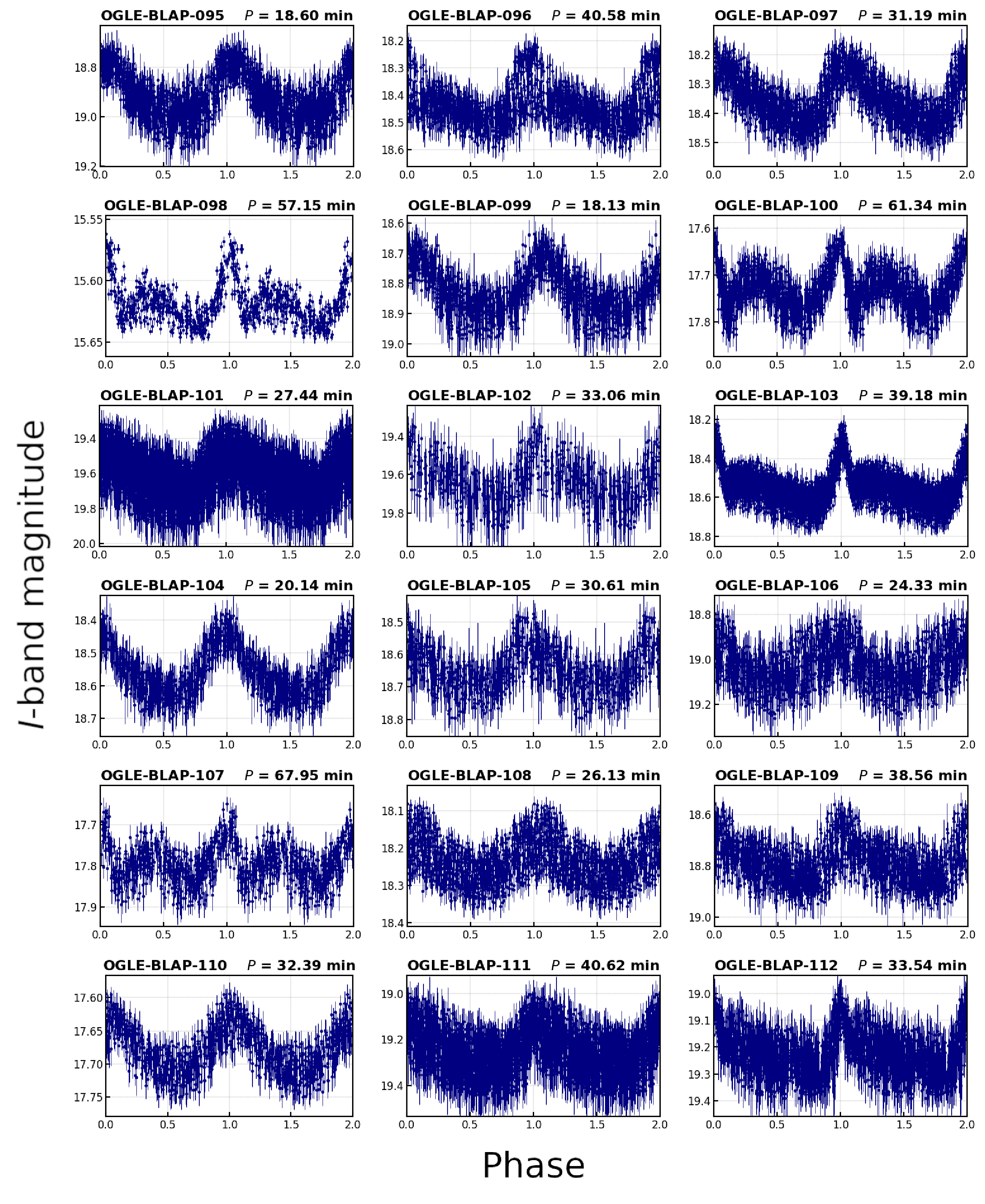}}
\FigCap{Phase-folded \textit{I}-band light curves of 18 (out of 90) new BLAPs detected in the OGLE inner Galactic bulge fields. For clarity, only a subset of epochs is shown: up to the first 2000 epochs when available, or fewer when period/amplitude variations are present.}
\end{figure}

\begin{figure}
\centerline{\includegraphics[width=13.71cm]{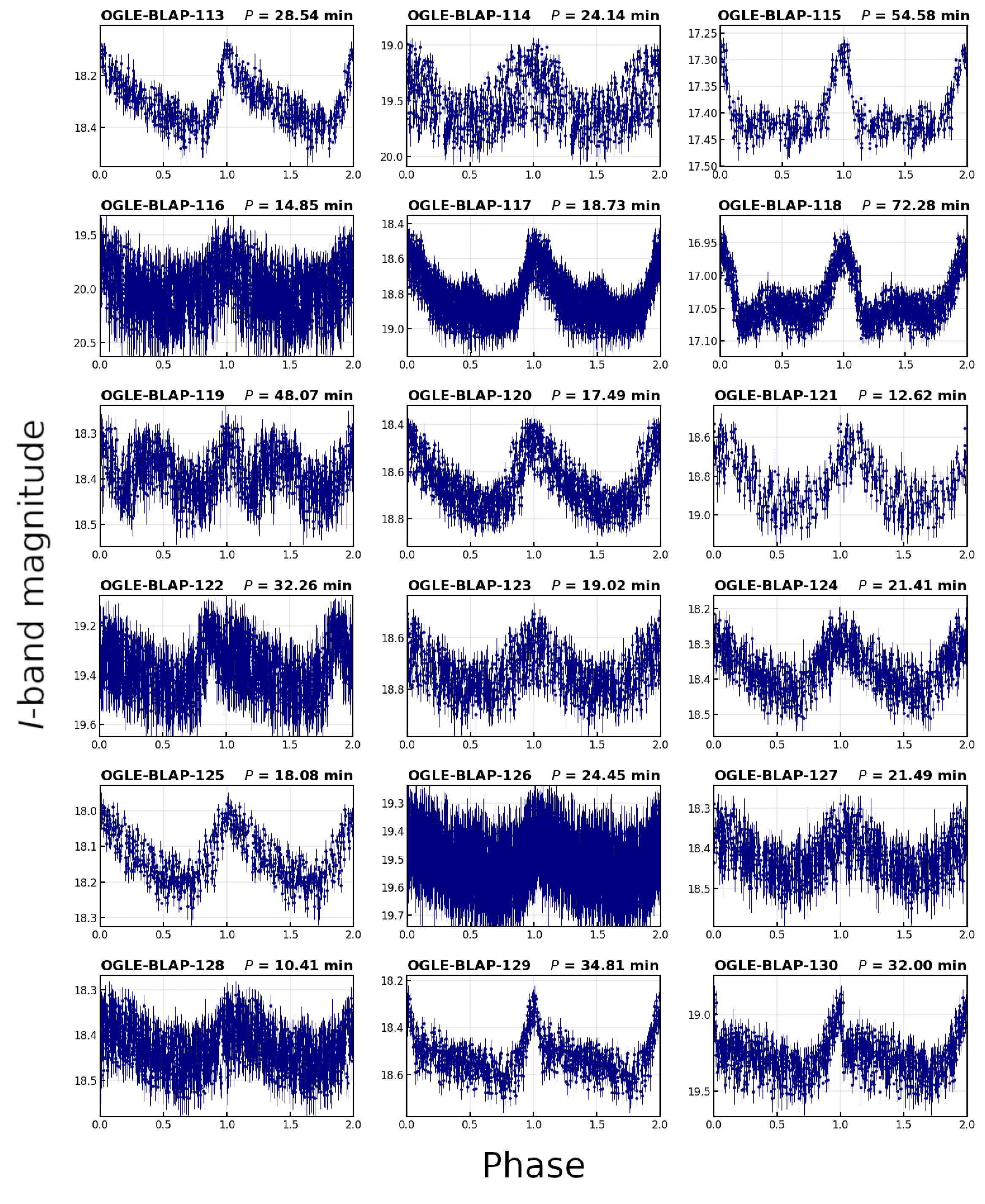}}
\FigCap{Fig. 6 (continued): Phase-folded \textit{I}-band light curves of another 18 BLAPs.}
\end{figure}

\begin{figure}
\centerline{\includegraphics[width=13.71cm]{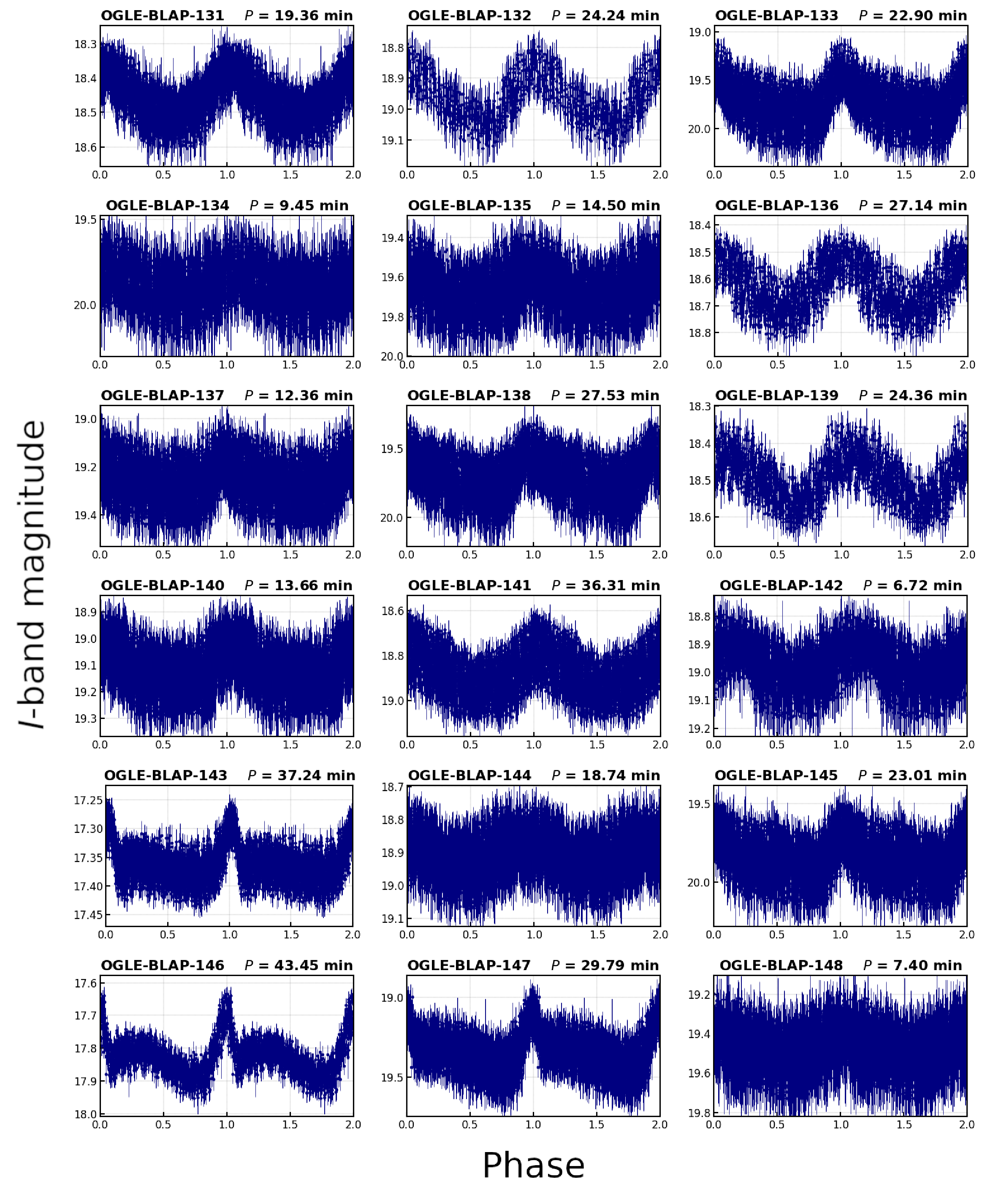}}
\FigCap{Fig. 6 (continued): Phase-folded \textit{I}-band light curves of another 18 BLAPs.}
\end{figure}

\begin{figure}
\centerline{\includegraphics[width=13.71cm]{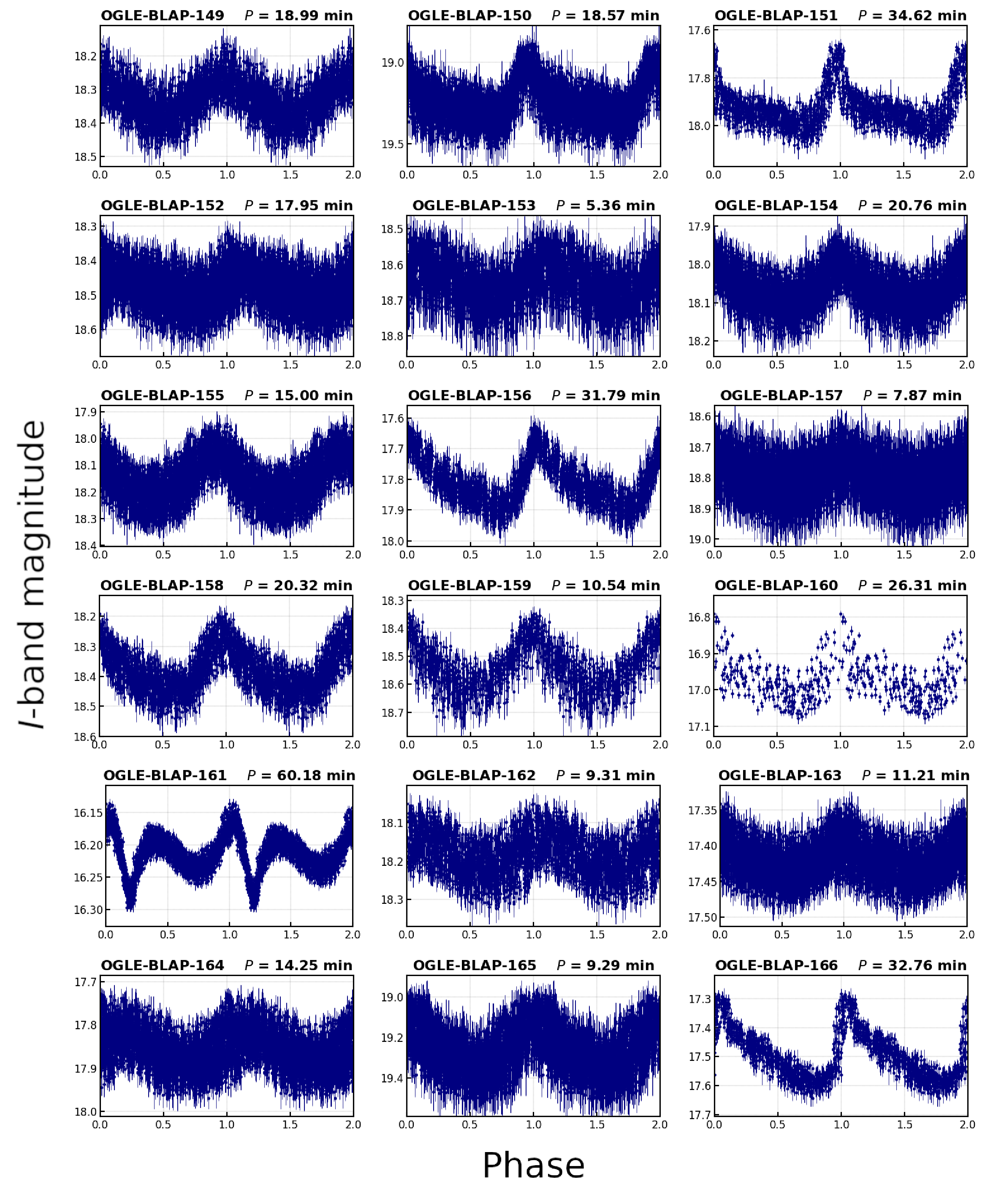}}
\FigCap{Fig. 6 (continued): Phase-folded \textit{I}-band light curves of another 18 BLAPs.}
\end{figure}

\begin{figure}
\centerline{\includegraphics[width=13.71cm]{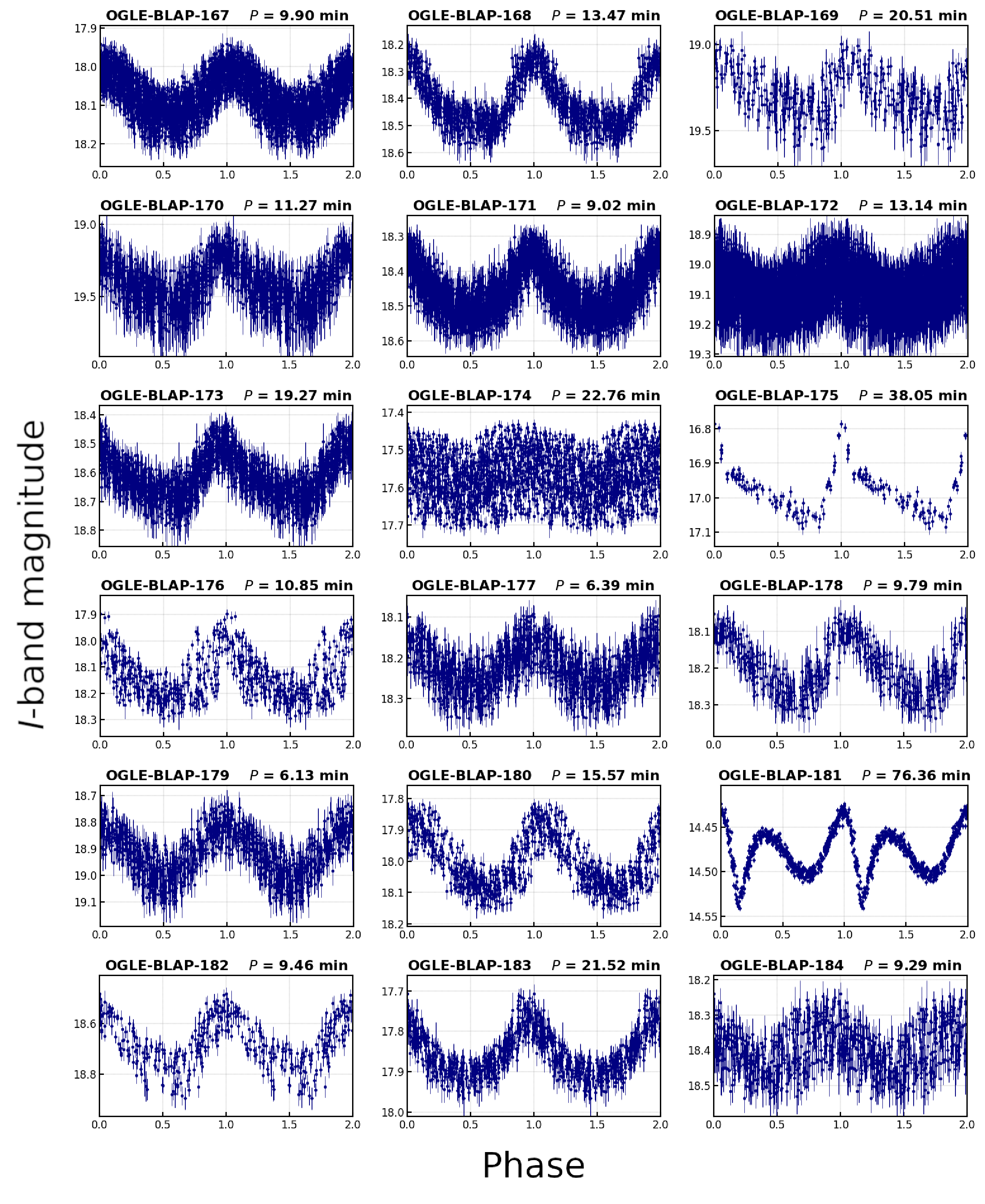}}
\FigCap{Fig. 6 (continued): Phase-folded \textit{I}-band light curves of another 18 BLAPs.}
\end{figure}
\section{Other Variables}

In this section, we present seven additional objects that may be of interest. These stars were identified in the same search that yielded the BLAPs, but their classification remains uncertain due to several ambiguous features. Basic observational parameters of these stars (sorted by OGLE-IV ID) are listed in Table~3. Fig.~11 shows phase-folded \textit{I}-band light curves together with \textit{I} \vs $(V-I)$ color-magnitude diagrams based on the OGLE-IV data. As a background, stars from the corresponding OGLE-IV subfields are plotted. Below, we provide a brief description of these variables.
\begin{table}[]
\scriptsize
\caption{List of other newly detected variables in the OGLE inner bulge fields}
\begin{tabular}{@{}lccrrrccccc@{}}
\toprule
\multicolumn{1}{c}{OGLE-IV ID}  & RA (J2000)       & Dec (J2000)       & \multicolumn{1}{c}{$l$}        & \multicolumn{1}{c}{$b$}        & Period    & \textit{I}       & $(V-I)$       & $A_I$   \\             
              & {[$^\circ$]}     & {[$^\circ$]}    & \multicolumn{1}{c}{[$^\circ$]}     & \multicolumn{1}{c}{[$^\circ$]}     & {[}min{]} & {[mag]}   & {[}mag{]}   & {[}mag{]}  \\ \midrule

BLG501.02.17335 & 268.39060 &$-$30.35137 &$-$0.32772 &$-$2.20753 & 7.39 & 16.332 & 0.223 & 0.029 \\
BLG535.12.23688$^{[*]}$ & 268.01866 &$-$32.39834 &$-$2.25503 &$-$2.97174 & 11.54 & 17.087 & 2.712 & 0.027 \\
BLG599.03.47429 & 268.08293 &$-$36.34094 &$-$5.63735 &$-$5.01222 & 4.25 & 18.211 & 0.862 & 0.085 \\
BLG611.12.191622$^{[*]}$ & 263.84336 &$-$27.21991 & 0.25748 & 2.82607 & 25.63 & 16.228 & 2.446 & 0.030\\
BLG621.06.29612 & 263.45985 &$-$22.67534 & 3.91058 & 5.56520 & 133.91 & 15.903 & 0.998 & 0.073 \\
BLG633.11.1021$^{[*]}$ & 266.61070 &$-$24.94708 & 3.50328 & 1.91784 & 31.00 & 17.239 & 2.418 & 0.058 \\
BLG633.24.104824 & 265.85694 &$-$24.51675 & 3.51380 & 2.72640 & 74.55 & 17.444 & 1.445 & 0.368 \\
\bottomrule 
\end{tabular}
\vspace{0.25em} 

\scriptsize{$[*]$ Likely to be a blended BLAP}
\end{table}

\textit{BLG599.03.47429} - This star is the shortest-period variable object detected in the entire history of the OGLE project, $P = 4.25$~min. Although the signal was found well above the adopted S/N threshold, the combination of its faintness and extremely short period (only about 2.5 times longer than the exposure time) prevents any detailed structure from being resolved in the light curve. It cannot be excluded that this is a short-period BLAP, as the observed amplitude of 0.085 mag is reduced by the time-averaging effect.  

\textit{BLG501.02.17335} - This star is among the shortest-period objects detected in our search. Multiple frequencies were identified in its light curve (see Sect.~6). It is also one of the bluest stars in our sample, with $(V-I) = 0.223$ mag. Data from Gaia DR3 (Gaia Collaboration 2021) confirm its hot nature, reporting $T_{\rm eff} \approx 18000 $~K and $\log g = 4.59$ dex, though its parallax is only $0.37 \pm 0.06$ mas. Owing to its low amplitude, this star is unlikely to be a BLAP. More likely it is a large-amplitude pulsating hot subdwarf. Further efforts, including spectroscopic observations, are required to resolve its nature.

Three additional objects clearly show light-curve shapes typical of BLAPs at their respective periods, but with amplitudes too low for classical BLAPs. These objects are \textit{BLG535.12.23688}, \textit{BLG611.12.191622}, and \textit{BLG633.11.1021}. The analysis of their positions in the color-magnitude diagram suggests that these stars are likely blended, as they show anomalously red colors for such short-period variables. We conclude that they are probably blended BLAPs, which explains the observed reduction in the amplitude.
\begin{figure}
\centerline{\includegraphics[width=10.5cm]{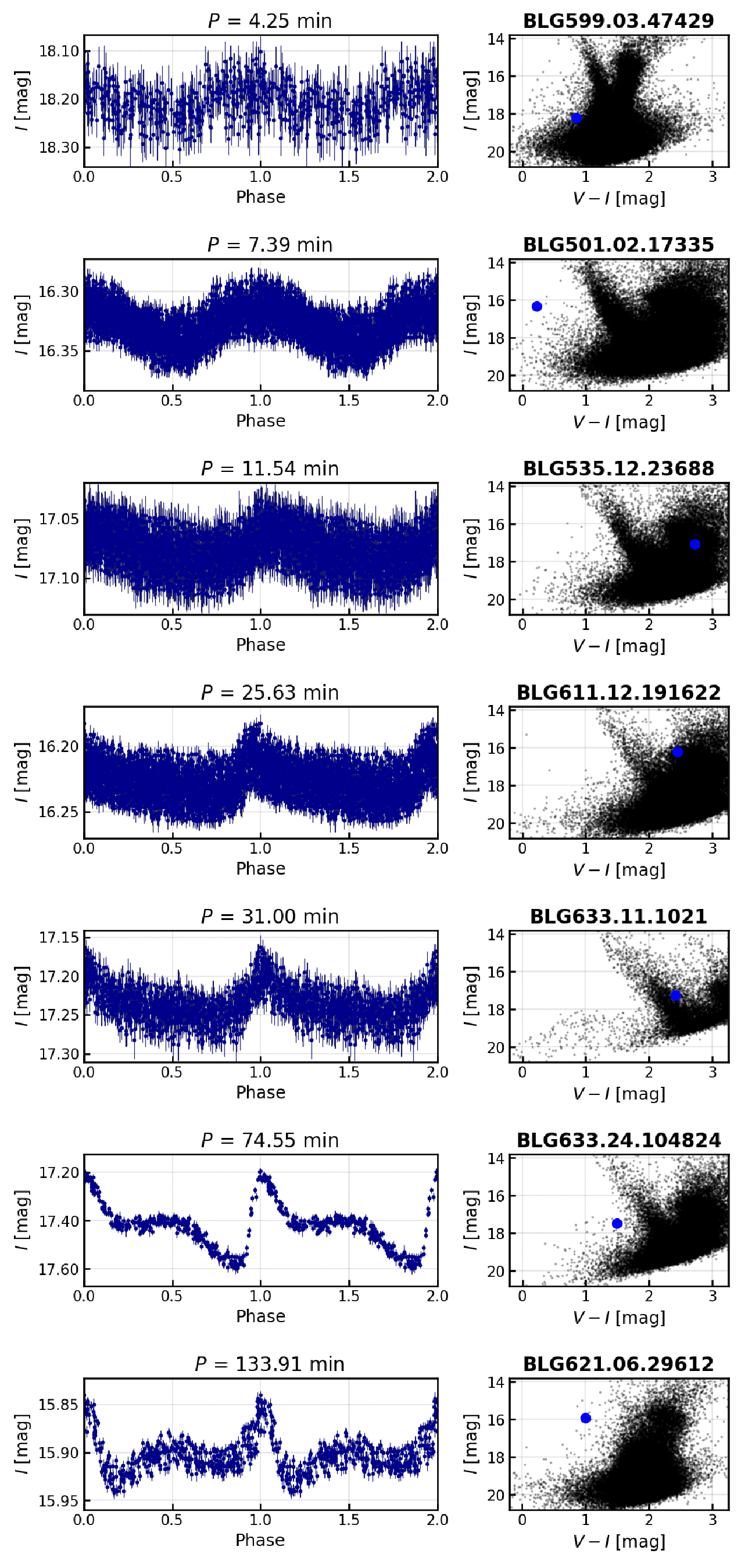}}
\FigCap{Left column: Phase-folded \textit{I}-band light curves of seven additional objects from the OGLE-IV inner Galactic bulge fields, Right column: \textit{I} vs. $(V-I)$ color magnitude diagrams with background stars taken from the corresponding OGLE-IV subfield.}
\end{figure}
\begin{figure}
\centerline{\includegraphics[width=13.7cm]{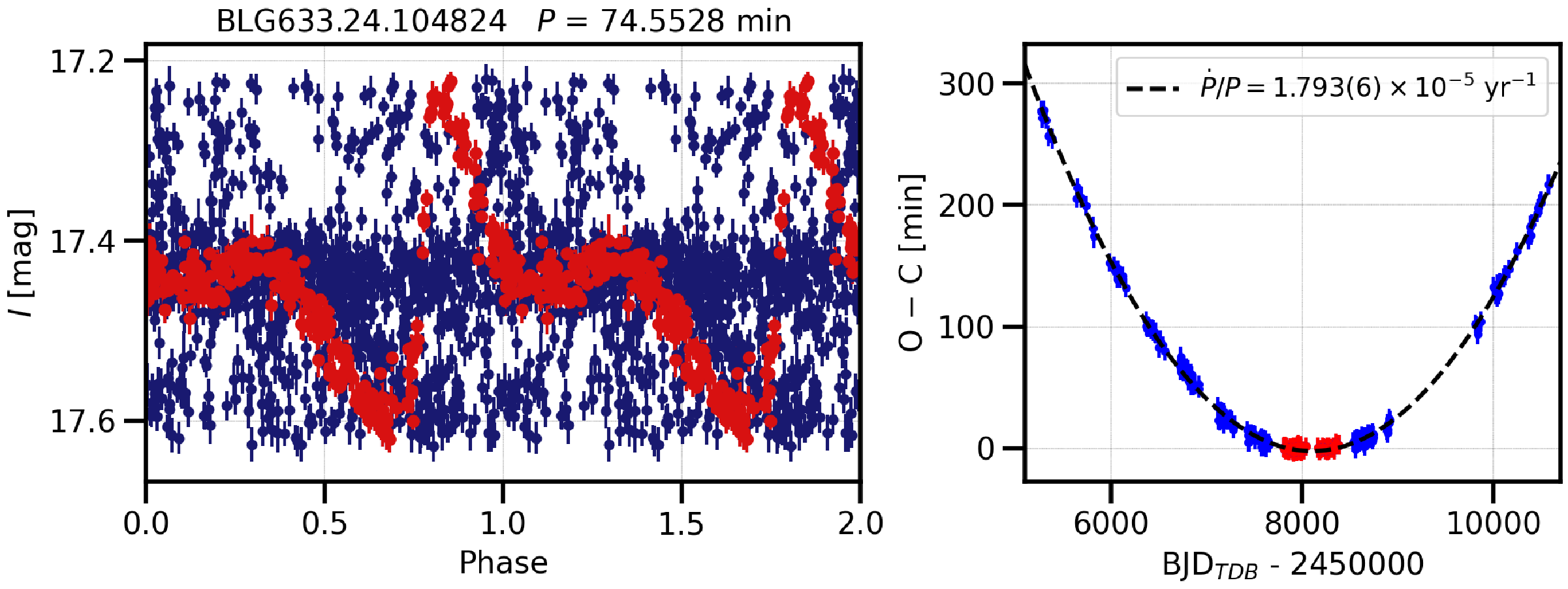}}
\FigCap{Left panel: Phase-folded light curve of object BLG633.24.104824, folded over two seasons (red) and entire observing period (blue). Right panel: O--C diagram for the same object, with quadratic function representing the best fit for the period change}
\end{figure}

An interesting case is the variable BLG633.24.104824, which at first glance appears to be a long-period BLAP with $P = 74.55$ min. A more detailed inspection, however, reveals an unusually high \textit{I}-band amplitude of 0.368 mag, which is atypical for long-period BLAPs, as these generally exhibit the lowest amplitudes among all BLAP stars. Another factor complicating the classification is the period change of this star. Fig.~12 shows the full light curve (blue), phase-folded using the best period found for two observing seasons (red), along with the O--C diagram and the corresponding best fit. The calculated times of maxima were determined by fitting a sixth-order Fourier series to the two observing seasons and then extrapolated over the full light curve, including both OGLE-III and OGLE-IV data. For the determination of observed maxima in the O--C analysis, we used the brightest 10\% of the data points, which is justified by the very sharp shape of the maximum in the light curve. For the maxima where the O--C exceeded half of the period, manual adjustments by that value were applied. A parabola was then fitted to the resulting diagram, yielding a rate consistent with a linear period change of $\dot{P} / P =  +1.793(6) \times 10^{-5}$ yr$^{-1}$. This value is approximately an order of magnitude higher than the largest period changes observed so far (TMTS-BLAP-1, OGLE-BLAP-030; Lin \etal 2023; Pietrukowicz \etal 2025a) and about two orders of magnitude larger than the typical rates previously observed in BLAPs. Given these unusual characteristics, we decided to withhold the classification of this object until additional data, particularly high-resolution spectroscopy, become available.

Object BLG621.06.29612, with a period of 133.91~min, represents the longest-period star reported in this study. In the color-magnitude diagram, it is located blueward of the main sequence. Although this star was previously identified as an ellipsoidal variable and is listed in the OCVS as OGLE-BLG-ELL-001829 (Soszy\'nski \etal 2016), it was independently discovered in our search. The light curve stood out among thousands of stars with comparable periods due to its pronounced asymmetry and the presence of a bump, reminiscent of those observed in long-period BLAPs. Preliminary studies suggested that this object may be the third known pulsating extreme helium star, similar to BX~Cir (Hill \etal 1981) and V652 Her (Kilkenny 1999), although confirmation of this classification is currently underway. Using the methodology outlined in Section 6, we derive a period-change rate of $\dot{P} / P = +6.3(1) \times 10^{-6}$ yr$^{-1}$, demonstrating a monotonically increasing period over the observed baseline. This trend is in contrast to that observed in BX Cir and V652 Her (Kilkenny \etal 2024), which exhibit decreasing periods.

\section{Periodicity Analysis}

In this section, we present the results of our periodicity analysis for the newly identified objects. For all variables reported in this study, the light curves were examined for the presence of additional frequencies. Table~4 summarizes the results for stars in which more than one frequency was detected. The listed frequencies were obtained after prewhitening the entire light curve with $f_1$ and its aliases. Subsequent frequencies are given together with their corresponding S/N ratio, where $f_1$ denotes the dominant frequency. In total, four BLAPs were found to exhibit additional periodicities, and there is evidence for a likely multimodality in the case of the variable BLG501.02.17335.

Fig.~13 shows a more detailed analysis of the additional frequencies, including their strength in individual observing seasons. For four BLAPs, OGLE-BLAP-132, OGLE-BLAP-144, OGLE-BLAP-155, and OGLE-BLAP-163, prewhitening with the dominant frequency and its aliases was performed season by season, followed by power spectrum analysis using \texttt{FNPEAKS}. The position of $f_1$ for each star is indicated by a red line. It is evident that both the strength and position of additional frequencies vary from season to season, although the effect of unequal numbers of data points in different seasons must also be taken into account. Whether these additional frequencies correspond to real pulsation modes remains a subject for further study. Daily aliases, inherent to ground-based observations, are also present. We note that since our sample of stars is, on average, relatively faint, and some of the detected objects approach the depth of the OGLE-IV survey (which implies photometric uncertainties above 0.1 mag), we cannot exclude the presence of additional frequencies that are too weak to be detected.

Eclipses have not been detected in any of the known BLAPs, including those reported in this study. Given the currently sizable sample, this absence strongly suggests that BLAPs do not reside in close binary systems but are instead more likely members of wide binaries, as further supported by O--C diagrams (e.g., Kim \etal 2025).
\begin{figure}
\centerline{\includegraphics[width=13.7cm]{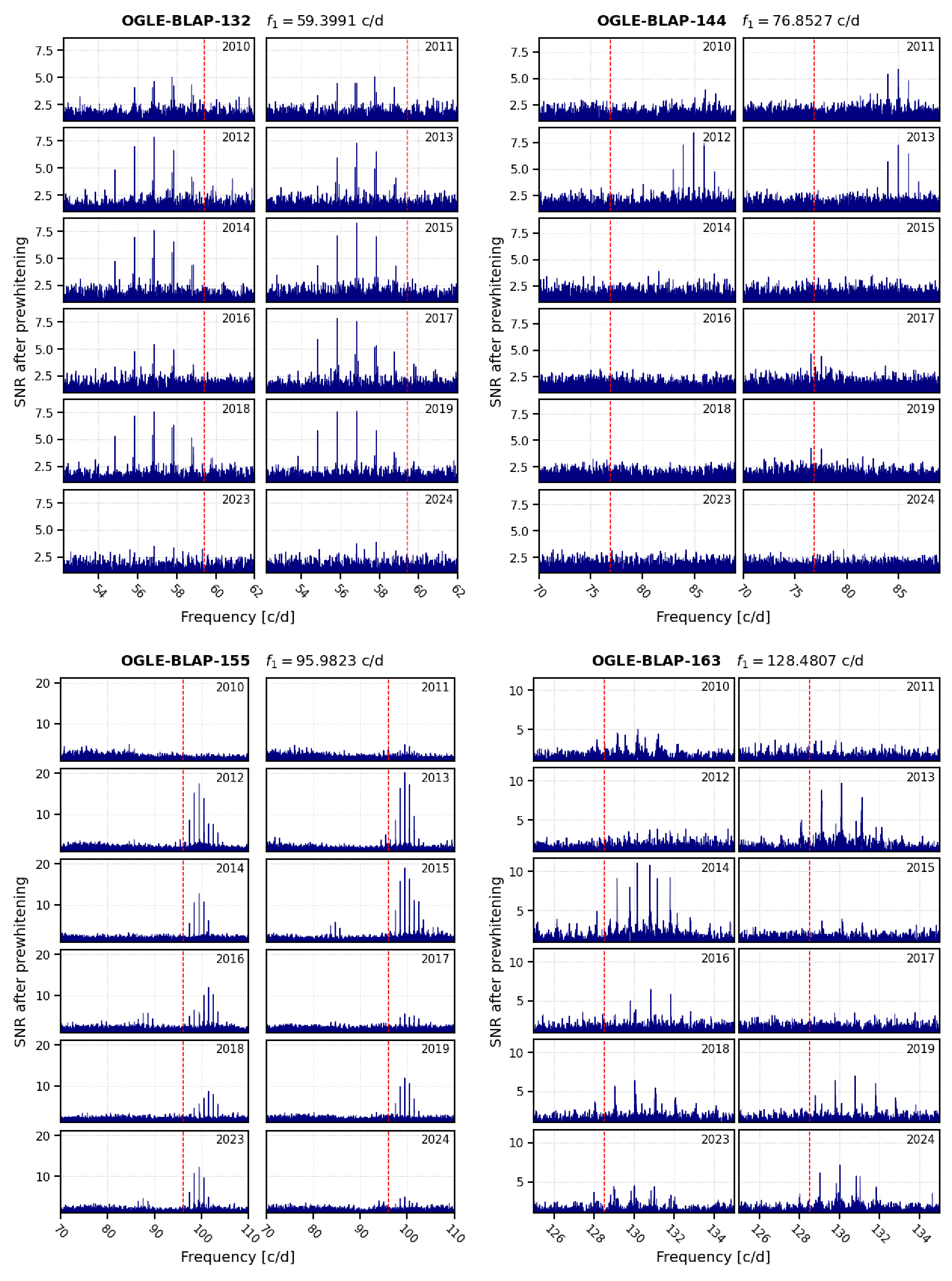}}
\FigCap{Periodogram analysis of four BLAPs exhibiting evidence for additional periodicities. For each object, the dominant frequency was identified from the full light curve and subsequently removed (prewhitened) from individual seasonal segments. The red vertical line indicates position of this subtracted frequency in each subplot. Every panel presents 12 periodograms corresponding to different observing seasons.}
\end{figure}
\begin{table}[ht]\
\footnotesize
\caption{Frequencies and S/N values for selected stars. }
\centering
\begin{tabular}{ccr}
\toprule
Star Name & Frequency [c/d] & S/N \\
\midrule
\multirow{3}{*}{OGLE-BLAP-132} 
  & $f_{1} = $ 59.3991 & 17.69 \\
  & $f_{2} = $ 56.8355 & 10.83 \\
  & $f_{3} = $ 57.7534 & 7.95 \\
\addlinespace

\multirow{2}{*}{OGLE-BLAP-144}
  & $f_{1} = $ 76.8527 & 14.04 \\
  & $f_{2} = $ 84.9748 & 8.60 \\
\addlinespace

\multirow{3}{*}{OGLE-BLAP-155}
  & $f_{1} = $ 95.9823 & 24.47 \\
  & $f_{2} = $ 99.4558 & 19.74 \\
  & $f_{3} = $ 84.6490 & 7.70 \\
\addlinespace

\multirow{3}{*}{OGLE-BLAP-163}
  & $f_{1} = $ 128.4807 & 11.16 \\
  & $f_{2} = $ 130.7878 & 7.53 \\
  & $f_{3} = $ 130.1630 & 7.01 \\
\addlinespace

\midrule

\multirow{4}{*}{BLG501.02.17335}
  & $f_{1} = $ 194.8481 & 32.35 \\
  & $f_{2} = $ 174.1376 & 18.06 \\
  & $f_{3} = $ 196.4946 & 13.42 \\
  & $f_{4} = $ 194.3480 & 13.02 \\
\bottomrule
\end{tabular}
\end{table}

To investigate the frequency changes over the years of observations, we follow the approach applied by Lin \etal (2023) in their analysis of TMTS-BLAP-1. For this purpose, we employed the normalized weighted wavelet Z-transform (WWZ, Foster 1996), which allows us to trace frequency variations in time across the entire sample of BLAPs reported in this work. The implementation was carried out using custom Python~3.8 software. Individual weight in the transform is defined as  
\[
w(t) = e^{-c(t - \tau)^2},
\]
where $w(t)$ denotes the weight assigned to each observation, $c$ is the parameter controlling the width of the time window, and $\tau$ is the central time point.

Because of the seasonal clustering of observations and the large variation in the number of available data points (ranging from a few hundred to several tens of thousands), the parameter $c$ was adjusted depending on the data density. The resulting $\sigma$ of the distribution across examined variables spans from several tens to about one hundred days. WWZ diagrams were then inspected at a resolution of 300 time bins and 300 frequency bins, across frequency ranges covering several orders of magnitude. In addition, we examined archival OGLE-III data as well as new measurements collected after the resumption of OGLE observations in August 2022.

The analysis of frequency changes within the sample revealed a variety of behaviors. Fig.~14 presents the WWZ results for stars in which systematic frequency changes were detected, representing the largest such values reported for BLAPs so far. The signal strength was normalized with respect to each bin and its adjacent bins. Irregularities on timescales of about every 300 days are artifacts caused by seasonal gaps in the observations. In the plots, the boundary between OGLE-III and OGLE-IV data is indicated. By tracking the frequency changes over the full observational baseline, we find that OGLE-BLAP-153 ($P = 5.36$~min) experiences a period change of $\dot{P} / P = +1.56(1) \times 10^{-5}$ yr$^{-1}$. The second object exhibiting a clear monotonic frequency modulation is OGLE-BLAP-176 ($P = 10.85$~min, OW-BLAP-1). Although this star was already discovered earlier (Ramsay \etal 2022), we report here for the first time such an extreme period change, of $\dot{P} / P = +1.58(1) \times 10^{-5}$ yr$^{-1}$.

Two other objects appear to exhibit possible periodic frequency modulation, as shown in Fig.~15. In OGLE-BLAP-136 ($P = 27.14$~min), we observe quasi-periodic modulation superimposed on a long-term trend. OGLE-BLAP-137 ($P = 12.36$~min) displays a cyclic frequency modulation, which could be explained by the light travel time effect (LTTE) in a binary system. Preliminary modeling suggests a wide binary with an orbital period of about 1150 days on an eccentric orbit ($e \approx 0.3$), although more detailed studies of this star are necessary.  

We conclude that most of the newly discovered BLAPs do not show period changes that can be unambiguously classified as either cyclic or linear. The conclusion from the remaining cases is that the majority of period variations in BLAPs appear to be random or stochastic, at a level of $\sim 10^{-7}$ yr$^{-1}$, consistent with earlier reports in the literature (Pietrukowicz \etal 2025a). Fig.~16 illustrates the WWZ results for two representative examples, showing, for instance, drops and discontinuities in pulsation frequency.
\begin{figure}
\centerline{\includegraphics[width=11cm]{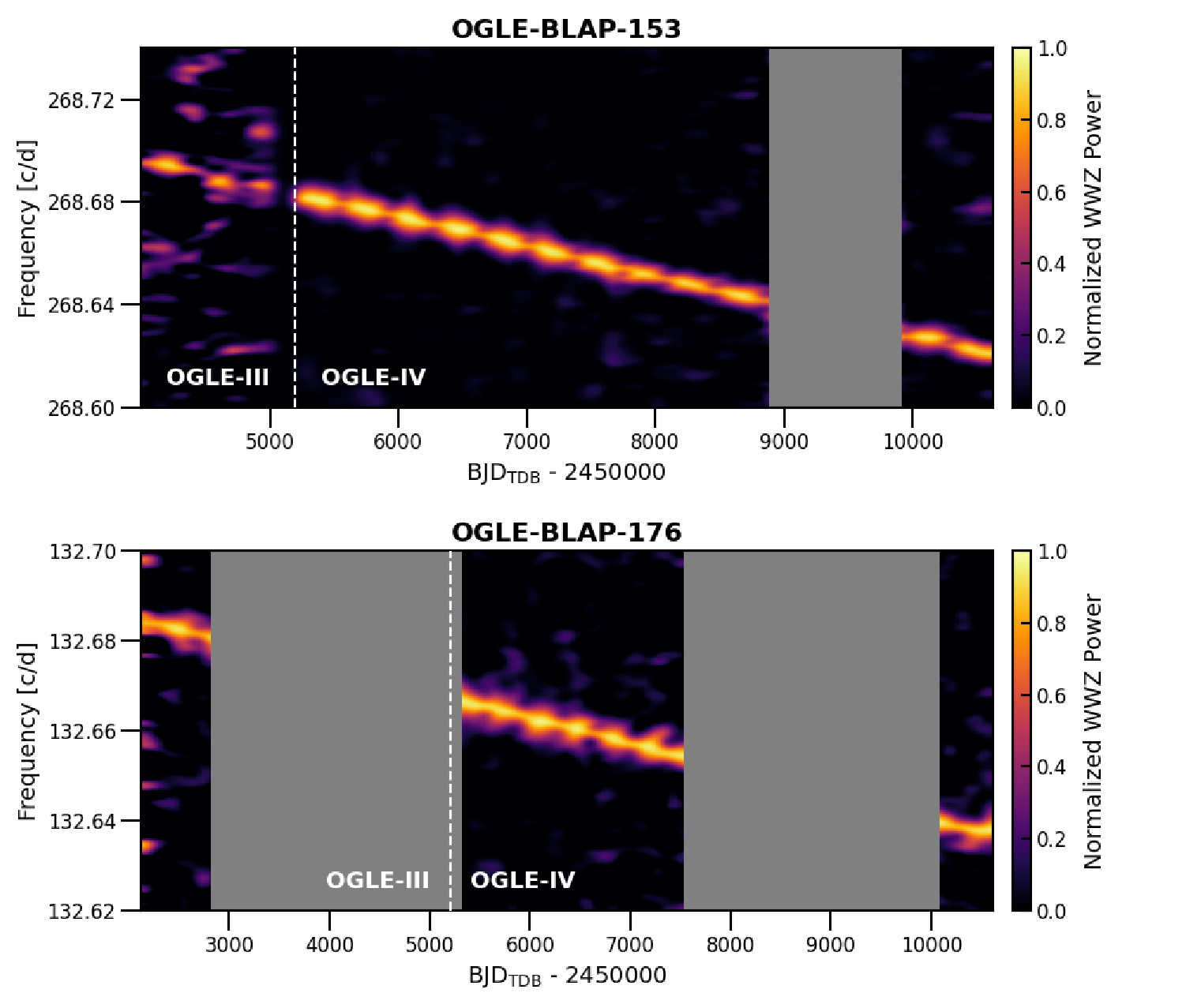}}
\FigCap{Normalized weighted wavelet Z-transform (WWZ) plots for BLAPs showing extreme frequency changes. Regions with sparse data are masked in gray.}
\end{figure}
\begin{figure}
\centerline{\includegraphics[width=11cm]{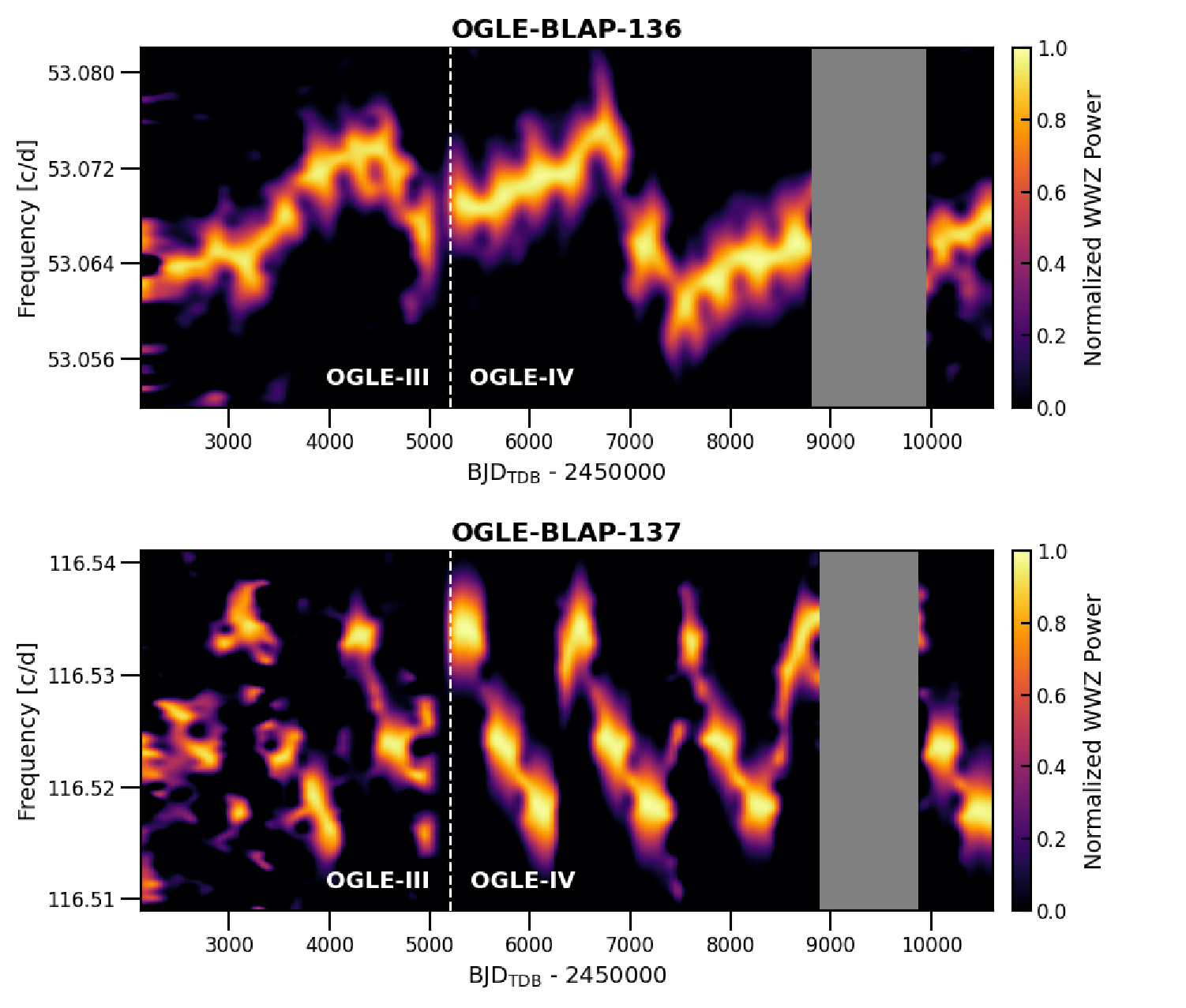}}
\FigCap{Normalized WWZ plots for BLAPs that show possible periodic frequency changes.}
\end{figure}
\begin{figure}
\centerline{\includegraphics[width=11cm]{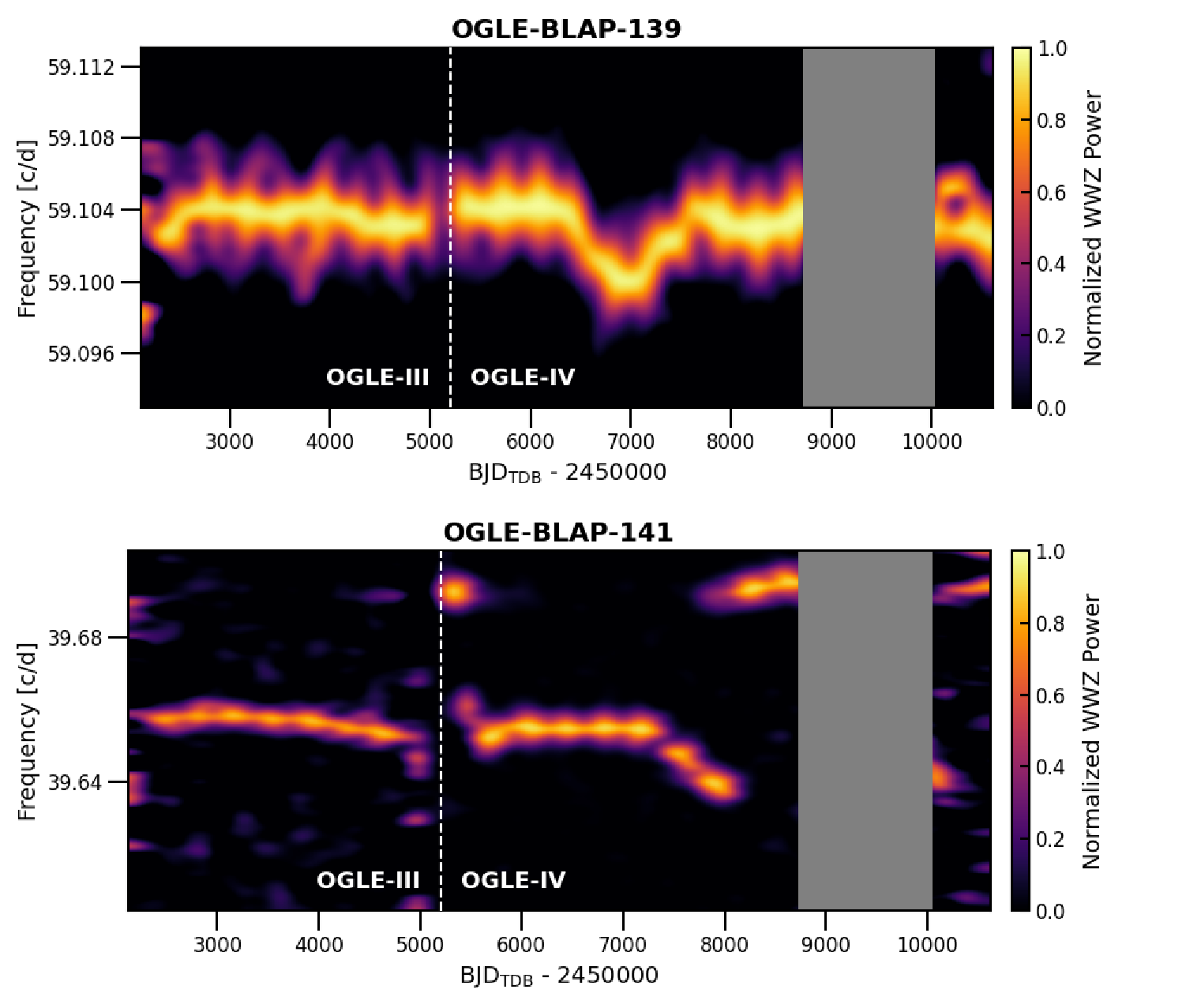}}
\FigCap{Normalized WWZ plots for two example BLAPs experiencing irregular behavior, including irregular frequency variations (upper panel) and discontinuities (lower panel).}
\end{figure}
\section{Completeness and Purity}

The overall completeness and purity of our BLAP collection are affected by several observational and instrumental factors. A primary limitation arises from the design of the OGLE-IV mosaic camera, which contains gaps between individual CCD detectors. These inter-chip regions reduce the absolute completeness of the catalog by several percent, since objects located within these areas were not monitored regularly or at all. A similar effect can occur with regard to the OGLE-IV fields. Although some fields partially overlap, there remain small regions, particularly in highly reddened parts of the bulge, that are not covered by any field. These effects were partially counteracted by searching the whole OGLE-III bulge database. 

Since no comparison catalog for BLAPs is currently available and these stars are intrinsically rare, completeness cannot be assessed in the traditional manner, i.e. by cross-matching with other surveys or by checking for double detections as in, for example, Soszy{\'n}ski \etal (2019). To estimate the completeness, we therefore carried out simulations assuming a representative light-curve shape, period, and amplitude. The results are shown in Fig.~17. Using the corrected OGLE-IV photometric uncertainties (Skowron \etal 2016), we simulated 100 artificial light curves for various combinations of the number of measurements and \textit{I}-band brightness, adopting the shape and amplitude of OGLE-BLAP-014. This star was chosen because it represents a typical member of the classical BLAP population. We then checked how many of these simulated light curves could be recovered using the methods described above. For a given set of mean \textit{I}-band magnitude and number of observations, percentage of recovered objects is color-coded. Seasonal clustering and daily aliasing were explicitly taken into account. The recovery fraction mirrors the behavior of the photometric error as a function of magnitude, as expected. Positions of OGLE BLAPs detected in the inner BLG fields are marked with white crosses. 

Combining these simulations with the distribution of the number of available measurements (shown in Fig.~18) allows us to estimate the completeness of our survey. It should be emphasized, however, that the simulations probe only the detectability of stars above a given S/N cutoff. For very faint stars, near the survey detection limit, reliable classification would likely fail due to smearing of the light curve. Together with the fact that the majority of stars in the surveyed fields have less than 1000 measurements (see Fig.~18), our conclusion is that the completeness for typical BLAPs is very high and close to 100\% for $I < 18.5$ mag, within the regions covered by OGLE-IV fields.

Several factors may contribute to the reduction of completeness, although quantifying their precise impact is difficult. The primary factor is a decrease in pulsation amplitude, which reduces the S/N and effectively shifts the detection limit toward brighter stars. The second factor are very short pulsation periods, which reduce the observed amplitude. However, in the case of OGLE data this effect becomes significant for periods only below about 10~min. Rapid period changes may also reduce the completeness by producing phase smearing in folded light curves. Nevertheless, this effect was partially mitigated by conducting additional searches on a season by season basis. An additional factor that may reduce the detection efficiency is source blending, which can lead to an underestimation of the observed amplitude. Blending is unlikely to pose a significant problem for bright stars, as the probability of a comparably luminous source being located along the same line of sight is relatively low; however, this assumption may not hold for some of the fainter objects. Another factor could be the non-sinusoidal, asymmetric shape of BLAP light curves, which reduces S/N, especially for stars with additional bumps. Since such stars are usually of higher luminosity, this effect is unlikely to play a major role. An open question remains whether some objects rejected as BLAPs, for example, those with low amplitudes or nearly sinusoidal light curves, may in fact belong to this class. Without a broader spectroscopic follow-up campaign, we cannot exclude that some lower amplitude BLAPs may have been missed, if such objects truly exist.

As for catalog purity, we expect it to be very high. While we cannot rule out single cases of contamination, we consider the purity to be essentially 100\% for BLAPs with typical sawtooth-shaped, high-amplitude light curves, given the absence of any known contaminants exhibiting this features in the relevant period range. For long-period BLAPs with highly asymmetric profiles, AM~CVn systems may represent a potential source of contamination. This possibility, however, is strongly disfavored by the spectroscopic confirmation for most of these stars and by the stability of their pulsation periods, amplitudes, and light-curve morphology. Another possible contaminant are intermediate polars, where variability on timescales comparable to BLAP pulsations may arise from white dwarf spin modulation. Yet this interpretation can also likely be excluded, since no additional longer periods are detected in the light curves, no X-ray counterparts are present in cross-matched catalogs, and no evidence for outbursts or long-term irregular trends has been found. Purity may be somewhat lower for stars with short pulsation periods ($P \lesssim 20$~min), since these BLAPs tend to show more sinusoidal light curves, have brightness near the detection limit, and could be mimicked by large-amplitude pulsating hot subdwarfs (e.g. Kupfer \etal 2021). No distinction was made between BLAPs and radially pulsating sdB stars, as photometry alone provides insufficient reliability for such classification. Additional contamination may arise from variable WDs of the ZZ Ceti and V777 Herculis types, which can pulsate in multiple modes and reach amplitudes comparable to those observed in BLAPs (C{\'o}rsico \etal 2019). Parallax and proper-motion measurements argue against this possibility for most objects; however, for the faintest sources these indicators are not sufficiently reliable.
\begin{figure}
\centerline{\includegraphics[width=13.7cm]{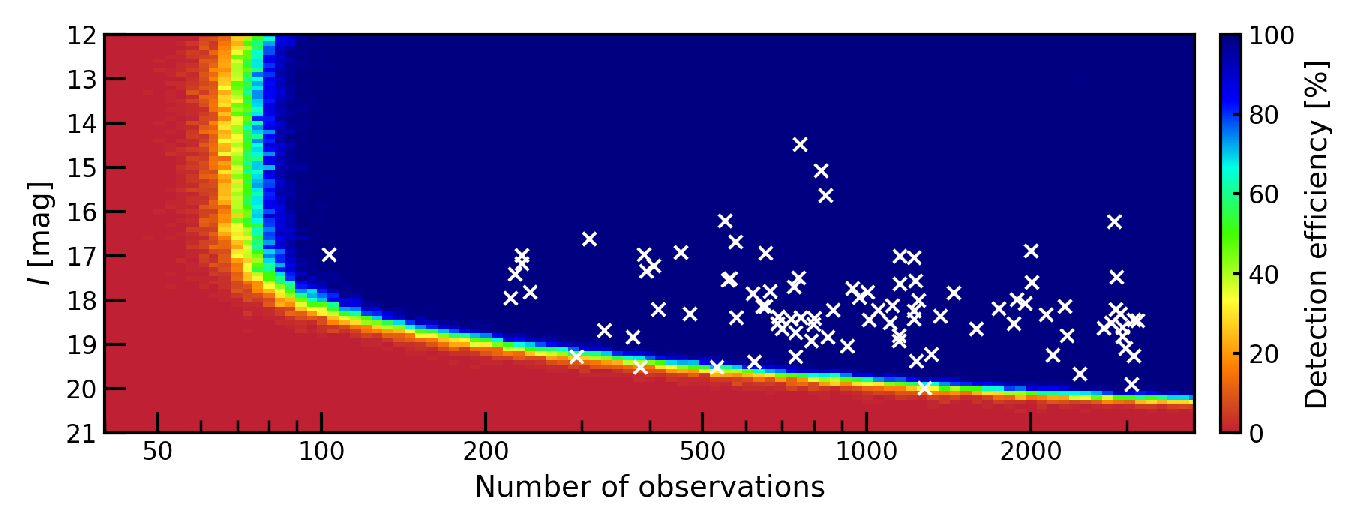}}
\FigCap{Simulation results for OGLE-BLAP-014, showing detection probability of object with the same \textit{I}-band amplitude and light curve shape. For each bin, 100 artificial light curves were simulated. BLAPs detected in the inner BLG fields are marked with white symbols.}
\end{figure}
Regarding a possible contamination by $\delta$~Sct or SX~Phe-type stars, based on comparative analysis and in-depth studies of pulsating stars that were not included in our collections (\eg Pietrukowicz \etal 2025b), we conclude that the periods at which these stars begin to appear with \textit{I}-band amplitudes above 0.1~mag are typically in the range of 40--45~min. Above this period threshold, all confirmed BLAPs exhibit clear asymmetries in their light-curve shapes, suggesting that distinguishing between these groups of variables on the basis of photometry alone is feasible. We therefore consider contamination by $\delta$~Sct or SX~Phe stars to be unlikely.
\begin{figure}
\centerline{\includegraphics[width=13.7cm]{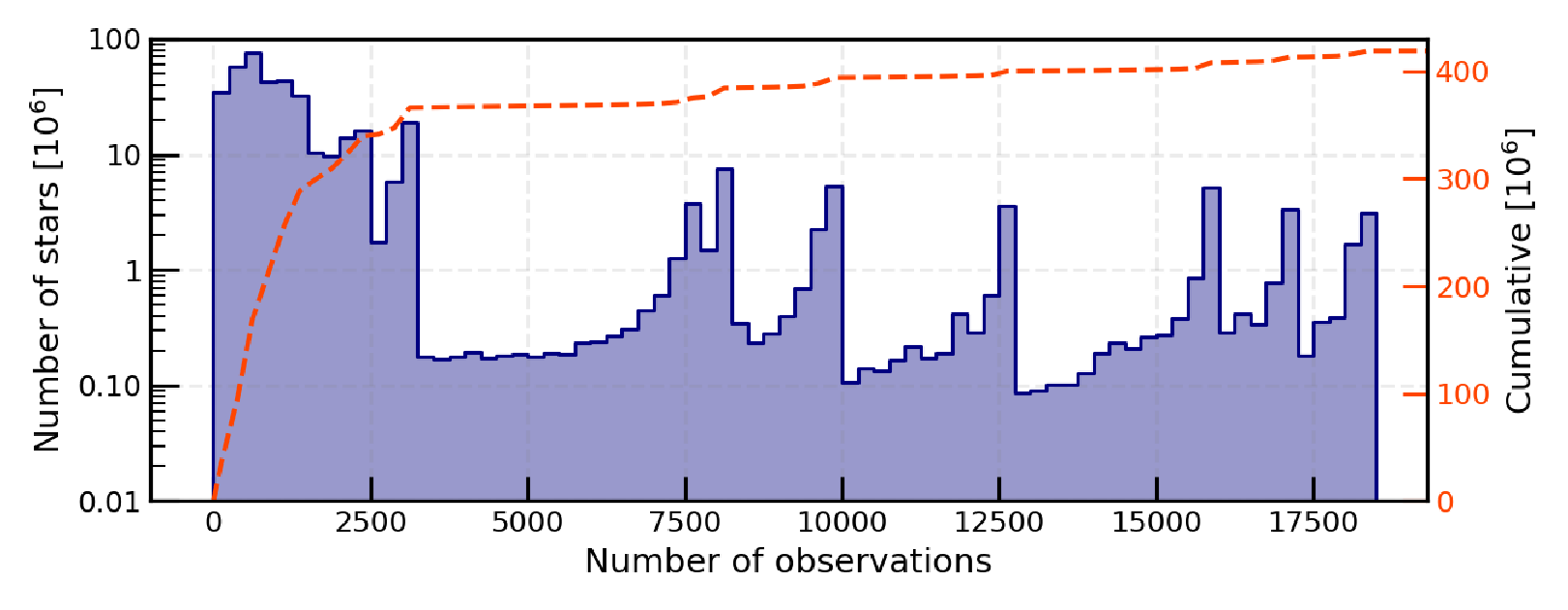}}
\FigCap{Distribution of stars as a function of the number of OGLE-IV epochs in the inner Galactic bulge fields. The red line represents the cumulative number of stars in millions.}
\end{figure}
\section{Summary}

By carefully analyzing OGLE-III and OGLE-IV data collected from 2001 to 2024, we searched over 400 million stars in the inner Galactic bulge for short-period variables. This search uncovered 87 new BLAPs, nearly doubling the total known population to around 200. Newly added BLAPs show pulsation periods from 5 to 76~min, with several objects further bridging the gap between classical BLAPs and high-gravity BLAPs.

The study provides full OGLE time-series photometry, maps the spatial distribution, and analyzes light-curve shapes, amplitudes, and period changes, with three reported stars showing unusually large period variations of the order of $10^{-5}$yr$^{-1}$ that may inform about their evolutionary status. Multiple frequencies were detected in four reported BLAPs. Completeness simulations indicate that our catalog is nearly complete for $I \lesssim 18.5$~mag, while the catalog purity for classical high-amplitude BLAPs is estimated to be 100\%. The dataset significantly extends the observed parameter space of BLAPs and is publicly available through OCVS, enabling further studies of their formation and evolution.

\Acknow{We thank all the OGLE observers for their contribution to the collection of the photometric data over the decades. P.P. has been supported by the Polish IDUB "Nowe Idee 3B" grant and "Microgrants" from the University of Warsaw, Poland. We used data from the European Space Agency (ESA) mission Gaia, processed by the Gaia Data Processing and Analysis Consortium (DPAC). Funding for the DPAC has been provided by national institutions, in particular the institutions participating in the Gaia Multilateral Agreement. This work is based on observations collected at the European
Southern Observatory under ESO programme 113.26U3 (PI: M.~J.~Mr\'oz).}

\end{document}